\documentclass[twocolumn,iop]{emulateapj}

\usepackage{natbib}
\usepackage{apjfonts}
\usepackage{subfigure}
\usepackage{url}
\usepackage[utf8]{inputenc}
\usepackage{fancyref}
\usepackage{hyperref}
\hypersetup{colorlinks=true,allcolors=blue,pdfborder={0 0 0}}

\shorttitle{Data remodelling and dynamical stability of HD~47366}
\shortauthors{J.~P. Marshall et al.}

\begin{document}

\title{Re-analysing the dynamical stability of the HD~47366 planetary system}

\author{J.~P. Marshall} 
\affil{Academia Sinica, Institute of Astronomy and Astrophysics, 11F Astronomy-Mathematics Building, NTU/AS campus,\\ No. 1, Section 4, Roosevelt Rd., Taipei 10617, Taiwan}

\author{R.~A. Wittenmyer, J. Horner, J. Clark,  M. W. Mengel}
\affil{University of Southern Queensland, Centre for Astrophysics, Toowoomba, QLD 4350, Australia}

\author{T.~C. Hinse}
\affil{Department of Astronomy and Space Science, Chungnam National University, Daejeon 34134, Republic of Korea}

\author{M.~T. Agnew}
\affil{Centre for Astrophysics and Supercomputing, Swinburne University of Technology, Hawthorn, Victoria 3122, Australia}

\author{S.~R. Kane}
\affil{Department of Earth Sciences, University of California, 900 University Avenue, Riverside, CA 92521, USA}

\begin{abstract}

Multi-planet systems around evolved stars are of interest to trace the evolution of planetary systems into the post-main sequence phase. HD 47366, an evolved intermediate mass star, hosts two giant planets on moderately eccentric orbits. Previous analysis of the planetary system has revealed that it is dynamically unstable on timescales much shorter than the stellar age unless the planets are trapped in mutual 2:1 mean motion resonance, inconsistent with the orbital solution presented in \cite{2016Sato} (hereafter: S16), or are moving on mutually retrograde orbits. Here we examine the orbital stability of the system presented in S16 using the $n$-body code {\sc Mercury} over a broad range of $a$--$e$ parameter space consistent with the observed radial velocities, assuming they are on co-planar orbits. Our analysis confirms that the system as proposed in S16 is not dynamically stable. We therefore undertake a thorough re-analysis of the available observational data for the HD 47366 system, through the Levenberg-Marquardt technique and confirmed by MCMC Bayesian methodology. Our re-analysis reveals an alternative, lower eccentricity fit that is vastly preferred over the highly eccentric orbital solution obtained from the nominal best-fit presented in S16. The new, improved dynamical simulation solution reveals the reduced eccentricity of the planetary orbits, shifting the HD 47366 system into the edge of a broad stability region, increasing our confidence that the planets are all that they seem to be. Our rigorous examination of the dynamical stability of HD 47366 stands as a cautionary tale in finding the global best-fit model.

\end{abstract}


\section{Introduction}
\label{sec:intro}

Multi-planet systems are important in revealing the influence of planet-planet interactions on the observed architectures and long-term stability of known planetary systems. Caution is needed, however - it is often the case that the best-fit solution for a given planetary system will place the planets therein on dynamically unfeasible orbits - ones that lead to collisions or ejections of those planets on timescales far, far shorter than the age of the system in which they reside \citep[e.g.][]{2011Horner,2012aWittenmyer,2012Horner}. In some cases, such dynamically unstable solutions are likely an indication that the observed behaviour of the star is driven by a process other than planetary companions \citep[e.g.][]{2012Horner,2013Wittenmyer,2013Horner,2014Horner}. In others, it can be a `red-flag' that points to the need for further observations in order to better constrain the proposed planetary orbits \citep[e.g.][]{2014Wittenmyer,2017bWittenmyer,2017Horner}.

The great majority of planet search programs have focused on `late-type' main-sequence stars - stars similar to, or less massive than the Sun \citep[e.g.][]{2006Jenkins, 2014Fischer, 2017Butler}. In the first two decades of the exoplanet era, when the radial velocity technique ruled supreme as a planet detection method, this was unsurprising. More massive main-sequence stars (O, B and A) have few spectral lines that can be used to determine radial velocities with the precision required for planet search work. On top of this, such stars are typically both active and rapid rotators - characteristics that greatly hinder the detection of small radial velocity signals. More recently, the balance has shifted somewhat, with the advent of large scale transit surveys for exoplanets such as the \textit{Kepler} mission \citep{2010Borucki,2016Coughlin} and ground-based programs such as Kilo-degree Extremely Little Telescope \citep[KELT;][]{2007Pepper}, Wide Angle Search for Planets \citep[WASP;][]{2006Pollacco}, and Hungarian Automated Telescope \citep[HAT;][]{2002Bakos}. However, our knowledge of the nature and frequency of planets around massive stars remains sparse, compared to our understanding of their less massive cousins \citep[e.g.][]{2010Bowler,2015Reffert, 2016Jones, 2017aWittenmyer}.

In particular, radial velocity measurements of `retired' intermediate 
mass stars are an excellent probe of the outcomes of planet formation 
around stars with masses 1.5 to 2.5~$M_{\odot}$, which are 
inaccessible to such surveys during their main-sequence lifetimes due to 
observational constraints. For this reason, several groups have begun 
radial velocity observations of `retired A-stars', whose cooler 
temperatures beget a suitable slew of absorption lines for analysis 
\citep[e.g.][]{2007Johnson}. Such work is complementary to the direct 
imaging surveys of young stars that preferentially examine intermediate 
mass stars \citep[e.g.][]{2013Janson,2016Durkan}, and provides a more 
complete picture of planet formation as a function of stellar mass 
\citep{2017Lannier}. Over the past few years, such studies have begun to 
bear fruit, with the discovery of an increasing number of planets 
orbiting these giant and sub-giant stars \citep[e.g.][]{Giant1,Giant2,Giant3}.

Beyond providing information on the occurrence of planets around more 
massive stars, planetary systems detected around evolved stars can also 
shed light on the manner in which planetary systems evolve as their 
stars age \citep[e.g.][]{2012Mustill,2013Mustill}. Of particular interest 
in this field are systems for which multiple planets can be detected. It 
seems likely that, as a star evolves off the main-sequence, the orbits 
and physical nature of its planets could be affected 
\citep[e.g.][]{2014Mustill}. Those planetary 
systems we find around such stars are the end product of that evolution 
process, and so it is critically important that we ensure that any such 
systems proposed are truly all they appear to be.

HD~47366 is an evolved, intermediate mass star. Two Jovian-mass 
companions were discovered by the Okayama and Xinglong Planet Search 
Programs \citep{2012Wang,2013Sato}. Dynamical modelling of these 
exoplanets determined that they were dynamically unstable unless they 
were either trapped in mutual 2:1 mean motion resonance, or were moving 
on mutually retrograde orbits {\citep[][hereafter S16]{2016Sato}}. The best-fit orbital 
solution proposed in S16 was well removed from the location 
of the 2:1 mean-motion resonance between the planets, making such a 
solution unlikely. Whilst mutually retrograde orbits often appear to 
offer a solution to such unstable scenarios 
\citep[e.g.][]{2011Horner,2012aWittenmyer,2014Horner}, they remain 
primarily of theoretical interest, as to obtain such orbits in practice 
without catastrophically destabilising the system requires an 
inordinately high degree of contrivance.

For this reason, the HD~47366 system is ripe for reanalysis, to 
determine whether it is truly dynamically feasible as proposed in the 
discovery work. Equally, if it were to prove unstable, it is interesting 
to consider whether an improved fit to the available data can be found 
for the planetary companions that would both describe the observational 
properties and maintain its dynamical stability over long periods.

Here we focus our extensive expertise with dynamical modelling of 
multiple (exo)planet systems on the case of HD~47366. In Section 
\ref{sec:obs}, we summarise the compiled radial velocities from 
literature sources used to model this system. In Sect. \ref{sec:ana} we 
present a dynamical analysis of the original S16 fit to the data, followed
by a complete refitting of the available velocities to determine a revised
architecture for the exoplanetary system. In Section 
\ref{sec:dis} we place the results of our stability analysis in context, 
comparing them to previous findings. Finally, in Section \ref{sec:con}, 
we summarise our findings and detail the conclusions of this work.

\section{Summary of system parameters and observations}
\label{sec:obs}

The physical properties (mass, radius, luminosity) adopted for HD~47366 
used in our simulations were taken from S16, for consistency in 
the modelling process. Other relevant values were taken from the literature.
A summary of relevant stellar properties are given in Table \ref{tab:star}. 
The planetary parameters, as proposed in S16, are presented in Table \ref{tab:planet}.

The radial velocities of HD~47366 are described fully in S16.  In 
brief, data were obtained from six instrumental configurations: (1) slit 
mode on the High Dispersion Echelle Spectrograph (HIDES) on the Okayama 
1.88m telescope (HIDES-S); (2) fibre mode on HIDES (HIDES-F); (3) the 
Coude Echelle Spectrograph on the 2.16m telescope at Xinglong Station 
with its old detector (CES-O); (4) the new detector on the Coude Echelle 
Spectrograph (CES-N); (5) the High Resolution Spectrograph on the 
Xinglong 2.16m (HRS); (6) the UCLES spectrograph on the 3.9m 
Anglo-Australian Telescope (AAT).

\begin{deluxetable}{lccc}
\tablewidth{0.45\textwidth}
\tablecolumns{4}
\tablecaption{Stellar parameters for HD~47366 as used in this work. \label{tab:star}}
\tablehead{
\colhead{Parameter}  & \colhead{Value}  & \colhead{Reference} \\
}
\startdata
Right Ascension (h m s) & 06 37 40.794 & 1 \\
Declination (d m s)     & -12 59 06.41 & \\
Distance (pc) & 12.5~$\pm$~0.42 & 2 \\
Spectral type & K1 III & 3 \\
$V$ (mag)     & 6.11~$\pm$~0.01 & 4 \\  
$T_{\rm eff}$ (K) & 4914~$\pm$~100 & 5 \\
$\log g$ (cm$^{2}$/s) & 3.10~$\pm$~0.15 & 5 \\
$R_{\star}$ ($R_{\odot}$) & 6.2~$\pm$~0.60 & 5 \\
$L_{\star}$ ($L_{\odot}$) & 24.5~$\pm$~3.2  & 5 \\
$M_{\star}$ ($M_{\odot}$) & 2.19~$\pm$~0.25 & 5 \\
Metallicity, [Fe/H] & -0.07~$\pm$~0.10 & 5 \\
Age (Gyr) & 0.94 & 5 \\
\enddata
\raggedright

\tablerefs{1. \cite{1997Perryman}; 2. \cite{2007vanLeeuwen}; 3. \cite{1988HoukSM}, 
4. \cite{2000Hog}, 5. \cite{2016Wittenmyer}. }

\end{deluxetable}

\begin{deluxetable}{lcc}
\tablewidth{0.45\textwidth}
\tablecolumns{3}
\tablecaption{Planetary parameters for HD~47366 from S16. \label{tab:planet}}
\tablehead{
\colhead{Parameter}  & \colhead{HD~47366 b}  & \colhead{HD~47366 c} \\
}
\startdata
$P$ (d)    & 363.3$^{+2.5}_{-2.4}$ & 684.7$^{+5.0}_{-4.9}$ \\
Mean anomaly (\degr)  &  288.7~$\pm$~75.3 & 93.5~$\pm$~35.4\\
$T_{\rm peri}$ (BJD-2450000) & 122$^{+71}_{-55}$ & 445$^{+55}_{-62}$ \\
$K$ (m/s) & 33.6$^{+3.6}_{-2.8}$ & 30.1$^{+2.1}_{-2.0}$ \\
$e$        & 0.089$^{+0.079}_{-0.060}$ & 0.278$^{+0.067}_{-0.094}$ \\
$\omega$ (\degr) & 100$^{+100}_{-71}$ & 132$^{+17}_{-20}$ \\
$m \sin i$ ($m_{\rm Jup}$) & 1.75$^{+0.20}_{-0.17}$ & 1.86$^{+0.16}_{-0.15}$ \\
$a$ (au)   & 1.214$^{+0.030}_{-0.029}$ & 1.853$^{+0.045}_{-0.045}$ \\ 
\enddata
\raggedright

\end{deluxetable}

\section{Analysis}
\label{sec:ana}

\subsection{Dynamical simulations of the S16 solution}
\label{sec:dyn_horner_1}

We apply two distinct methods to survey the orbital phrase space around 
the best-fit solution for the HD~47366 planetary system provided in 
S16. The first technique provides the dynamical context of 
the solution, yielding dynamical maps of the orbital element space 
around the best-fit orbit for the planet with the least constrained 
orbital elements \citep[see e.g.][]{2010Marshall,2012bWittenmyer,2014Wittenmyer}. The second 
technique, which we first deployed in \cite{2017bWittenmyer}, simulates a 
large number of planet pairs distributed around the best-fit solution in 
$\chi^2$ space. The cloud of such solutions maps the stability of the 
system as a function of the goodness of fit to the observational data.  

The contextual method developed to analyse such systems goes as follows. 
We create dynamical maps that show the context of the orbital solutions 
proposed using the $N$-body code {\sc Mercury} \citep{1999Chambers}. To 
do this we run a large number (typically 126,075) of individual 
realisations of the planetary system in question, using a different 
initial set of orbital elements for the planet with the least 
constrained orbit (typically the outermost) in each realisation. Those 
solutions are generated in a hypercubic grid, centred on the best-fit 
solution. We then follow the evolution of the planets through simulation 
for a period of 100 million years, or until they either collide with one 
another, are ejected from the system, or collide with the central body.

In the case of HD~47366, the initial orbital parameters of the inner 
planet, HD~47366 b, were held fixed. {Our motivation for holding 
planet b’s parameters fixed were that the parameters of planet b are 
the better constrained of the pair, so by varying the less well-defined 
planet we survey a larger part of parameter space for the system’s 
potential properties. If we instead opt to hold planet c fixed, with its
slightly higher orbital eccentricity, we should expect a lower overall 
likelihood of a stable system configuration being found.} We therefore
tested realisations of the outer 
planet, HD~47366 c, incrementally adjusting the values of semi-major 
axis $a$, orbital eccentricity $e$, $\omega$ and mean anomaly, to probe 
a $\pm$3-$\sigma$ range around its best fit orbital parameters. 
To cover the 3-$\sigma$ parameter space we test 41 unique values of $a$ and $e$, 
i.e. at each point in semi-major axis space, we test 41 unique values of 
orbital eccentricity. For each of those locations in $a$--$e$ space, 
we tested 15 unique values of $\omega$, with five unique values of 
mean-anomaly tested for each unique $\omega$ examined. This gave a grid 
of 41 $\times$ 41 $\times$ 15 $\times$ 5 = 126,075 simulations.

We have assumed that the two planets are on co-planar orbits. This assumption 
is based on knowledge of the architectures of known multiple systems \citep{2011Lissauer,2012FangMargot}. 
It also provides a limiting case of maximum potential stability for the system 
(i.e. we are looking to maximise the opportunity for the system to yield dynamically 
feasible solutions that do not require mutual retrograde motion). We have also used the orbital parameters of the inner 
planet fixed at their best-fit values. This tacitly assumes that the inner planet 
has the better constrained orbit of the pair because it has the shorter orbital 
period and was thus better sampled by the radial velocity observations.

In addition to the contextual maps for the system, we performed an 
additional 126,075 simulations of potential architectures for the two 
planets involved, following the methodology laid out in 
\cite{2017bWittenmyer}, {also using {\sc mercury} for the $n$-body dynamical simulations}.
From the MCMC chain obtained by our refitting 
procedure (Section 3.2), we populated three ``annuli'' in $\chi^2$ space 
corresponding to the ranges 0 to 1-$\sigma$, 1 to 2-$\sigma$, and 2 to 3-$\sigma$ 
from the best fit.  Each annulus contained 42,025 solutions drawn from 
the MCMC chain ($10^7$ iterations).  The innermost annulus was drawn 
from the lowest 68.3~\% of all $\chi^2$ values, the middle annulus 
contained the next best 27.2~\% of values, and the outer annulus 
contained the worst 4.5~\% of solutions (i.e. those falling 2- to 3$-\sigma$ 
away from the best fit).

\begin{figure}
    \centering
    \includegraphics[width=0.35\textwidth,angle=270,trim={3cm 4cm 3cm 4cm},clip]{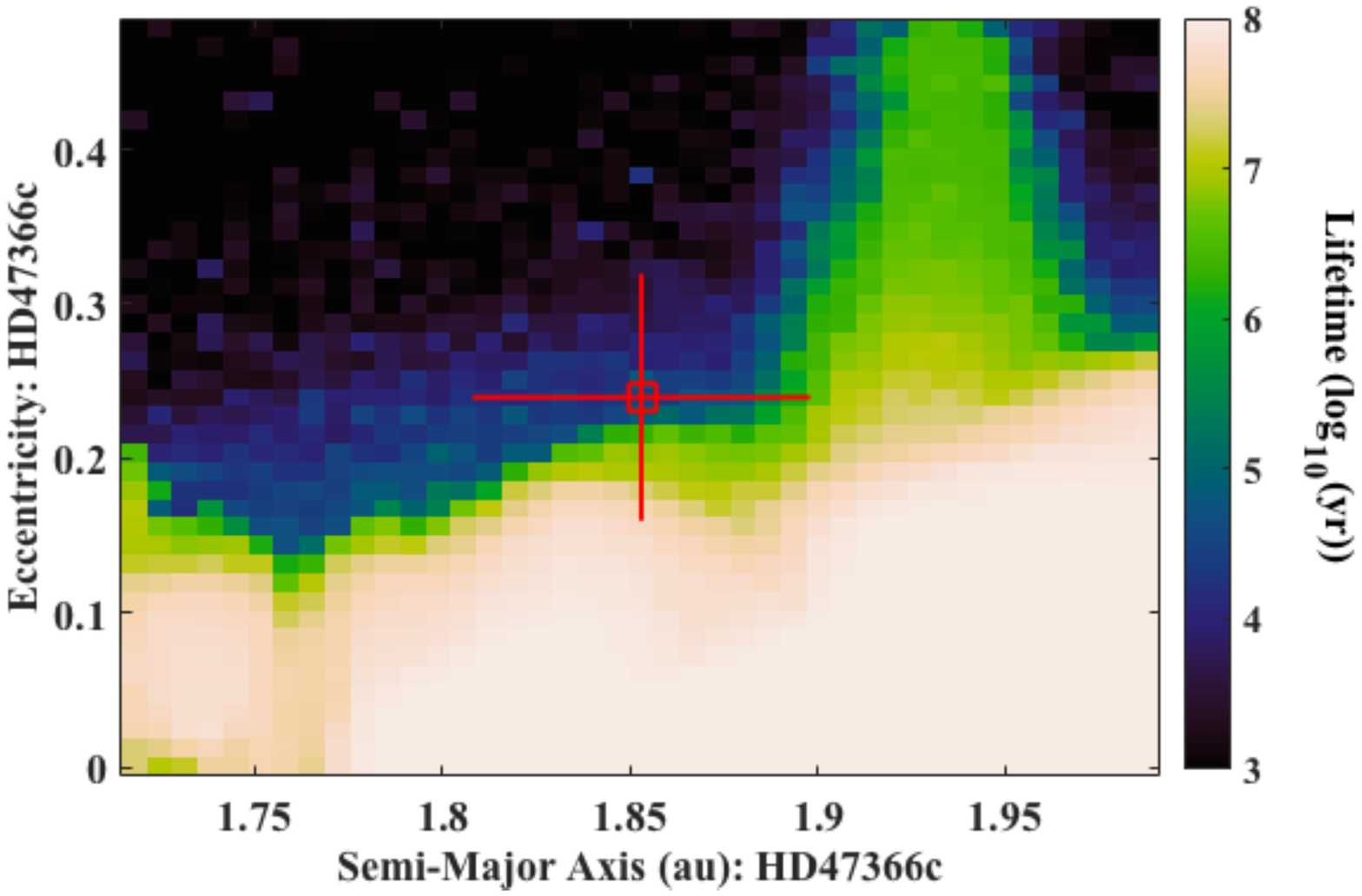}
    \caption{The dynamical stability of the S16 solution for the orbits of the two planets around HD~47366, as a function of the semi-major axis, $a$, and eccentricity, $e$, of HD~47366c. The best-fit solution for the system is marked by the hollow red box, with the published 1-$\sigma$ uncertainties on that solution denoted by the horizontal and vertical red lines that radiate from within. The lifetime shown at each location in the dynamical map is the mean of a total of 75 trials for that particular combination of $a$ and $e$. The best-fit solution for the system lies in a region of strong dynamical instability. In order to be dynamically stable, the orbit of HD~47366 c must be markedly less eccentric than was proposed in the discovery work. \label{fig:SatoContext}}
\end{figure} 

\begin{figure*}
    \includegraphics[width=0.5\textwidth]{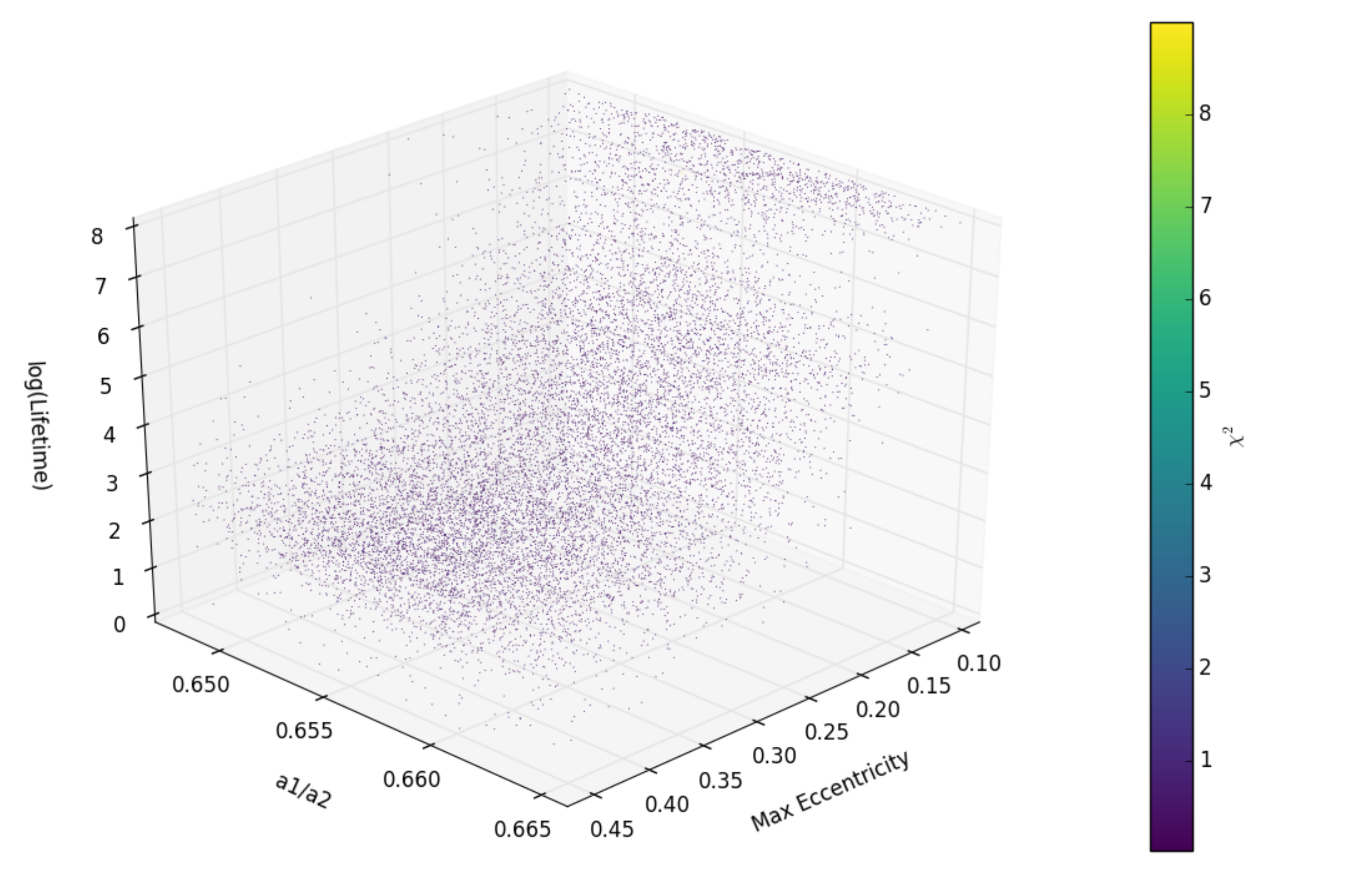}
	\includegraphics[width=0.5\textwidth]{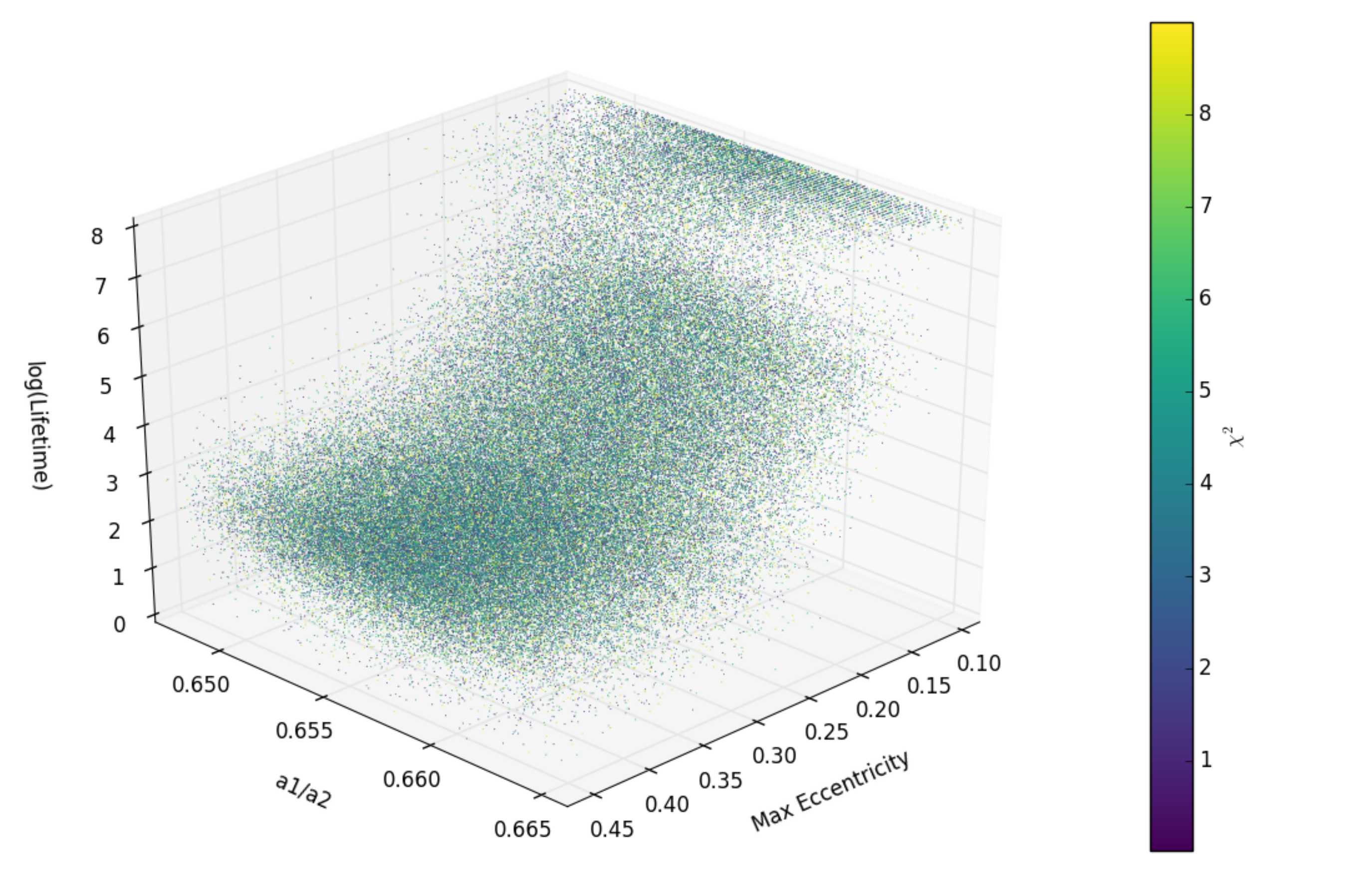}
    \includegraphics[width=0.5\textwidth]{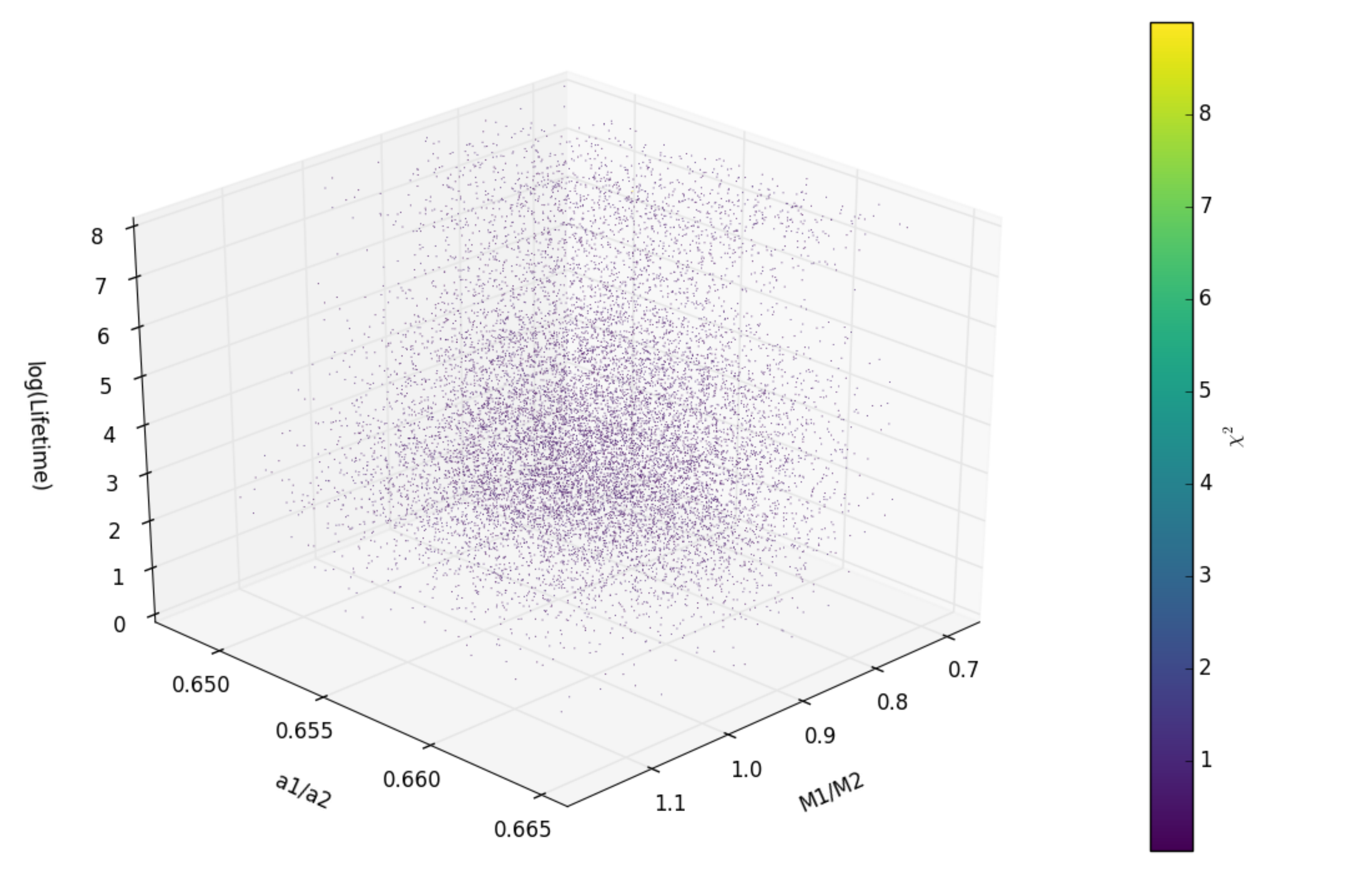}
	\includegraphics[width=0.5\textwidth]{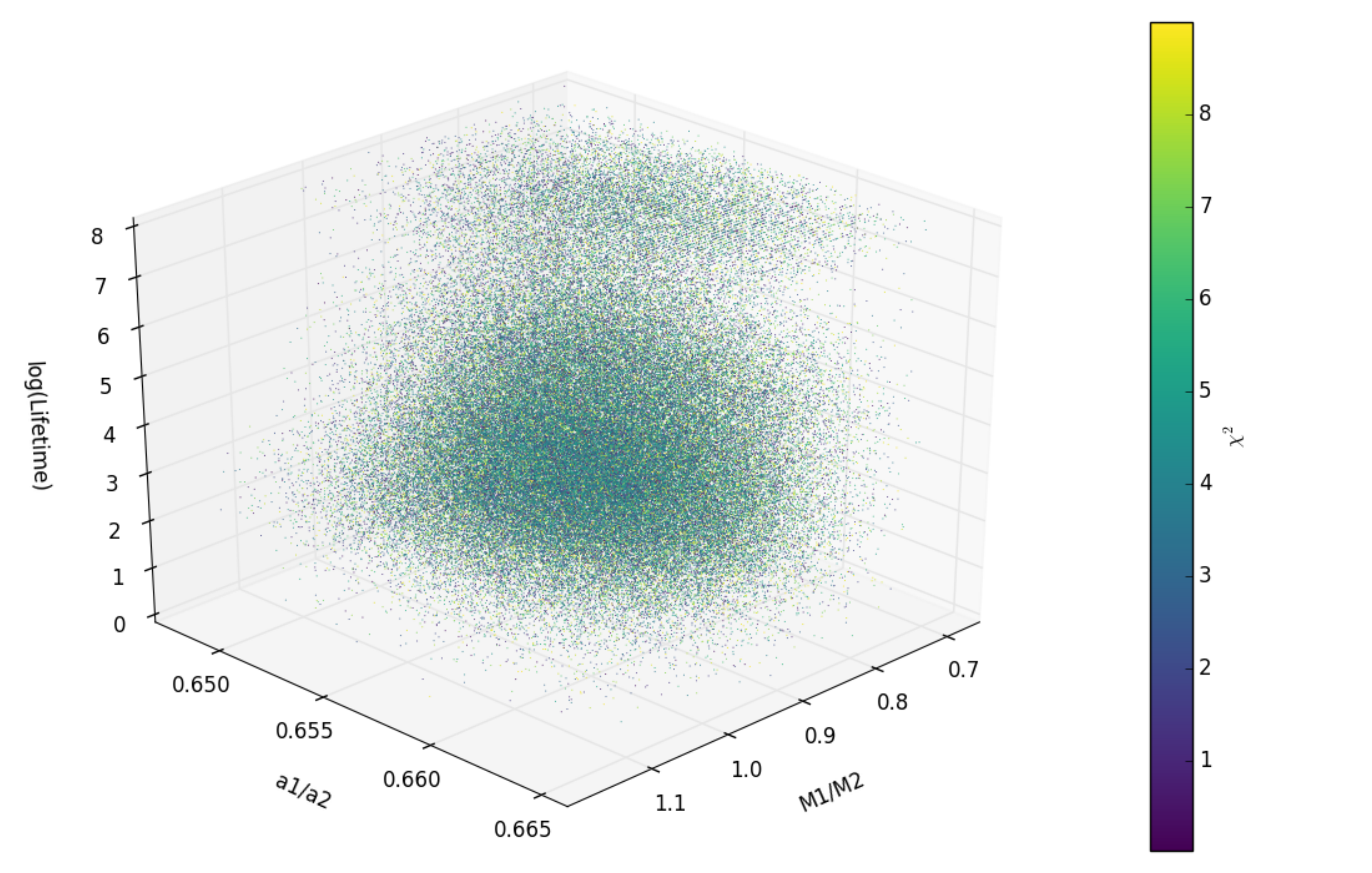}
\caption{The dynamical stability of S16's previously published solution for the HD47366 planets, as a function of the largest initial eccentricity fit to HD47366b and c, and the ratio of their orbital semi-major axes. The colour bar shows the goodness of fit of each solution tested, with the left plot showing only those results within 1-$\sigma$ of the best-fit case, and the right plot showing all solutions tested that fell within 3-$\sigma$ of that scenario. In the online version of this paper, the same plot is available in animated format. {The animated figure lasts 40~s, and shows the 1-$\sigma$ and 3-$\sigma$ distribution of points (equivalent to left and right panels on each line) from a changing perspective rotating around the $z$-axis (log(Lifetime)). These animations help illustrate the regions of parameter space that are more dynamically stable.}\label{SatoPairs}}
\end{figure*}

The results for our contextual simulations can be seen in Figure~\ref{fig:SatoContext}. It is immediately apparent that the best-fit solution lies in an a region of significant instability, and that the stable regions lie more beyond that bounded by the published 1-$\sigma$ uncertainties on the solution. Typically, stable solutions require an eccentricity for HD~47366 c below $\sim 0.2$. A region of moderate stability extends to high eccentricity at $a \sim 1.94$ au, the location of the mutual 2:1 mean-motion resonance between the two planets. However, at the eccentricity of the nominal best-fit solution, this region still only offers sufficient dynamical protection to yield mean lifetimes of order one million years. The results of our simulations of planet-pairs around the best-fit solution from S16 are shown in Figure~\ref{SatoPairs}.

These results are complemented by those shown in Figure~\ref{SatoPairs}, which presents the outcomes of our simulations of planetary solutions that fit the observational data to a given level of precision. In that figure, the left-hand panels show the lifetimes of simulations for solutions that provided a fit to the observational data within 1-$\sigma$ of the best-fit, whilst the right-hand panels present solutions that fell within 3-$\sigma$ of the best-fit outcome. It is immediately apparent that very few of the systems tested proved dynamically stable on multi-million year timescales. Those that did were all found in scenarios that featured orbital eccentricities of less than 0.1 for both planets in the system. This is not a great surprise, as reducing the eccentricity of the orbits of a given planet pair whilst keeping all other variables constant will increase the distance between the planets at their closest approach, and therefore lessen the impact of mutual encounters on the system's long term stability. It should also be noted that no solutions were found that placed the two planets in mutual mean-motion resonance - ruling out the mutual 2:1 mean-motion resonance as a source of stability for the system.

\begin{figure*}
    \centering
    \includegraphics[width=0.25\textwidth,trim={3cm 5cm 3cm 5cm},clip]{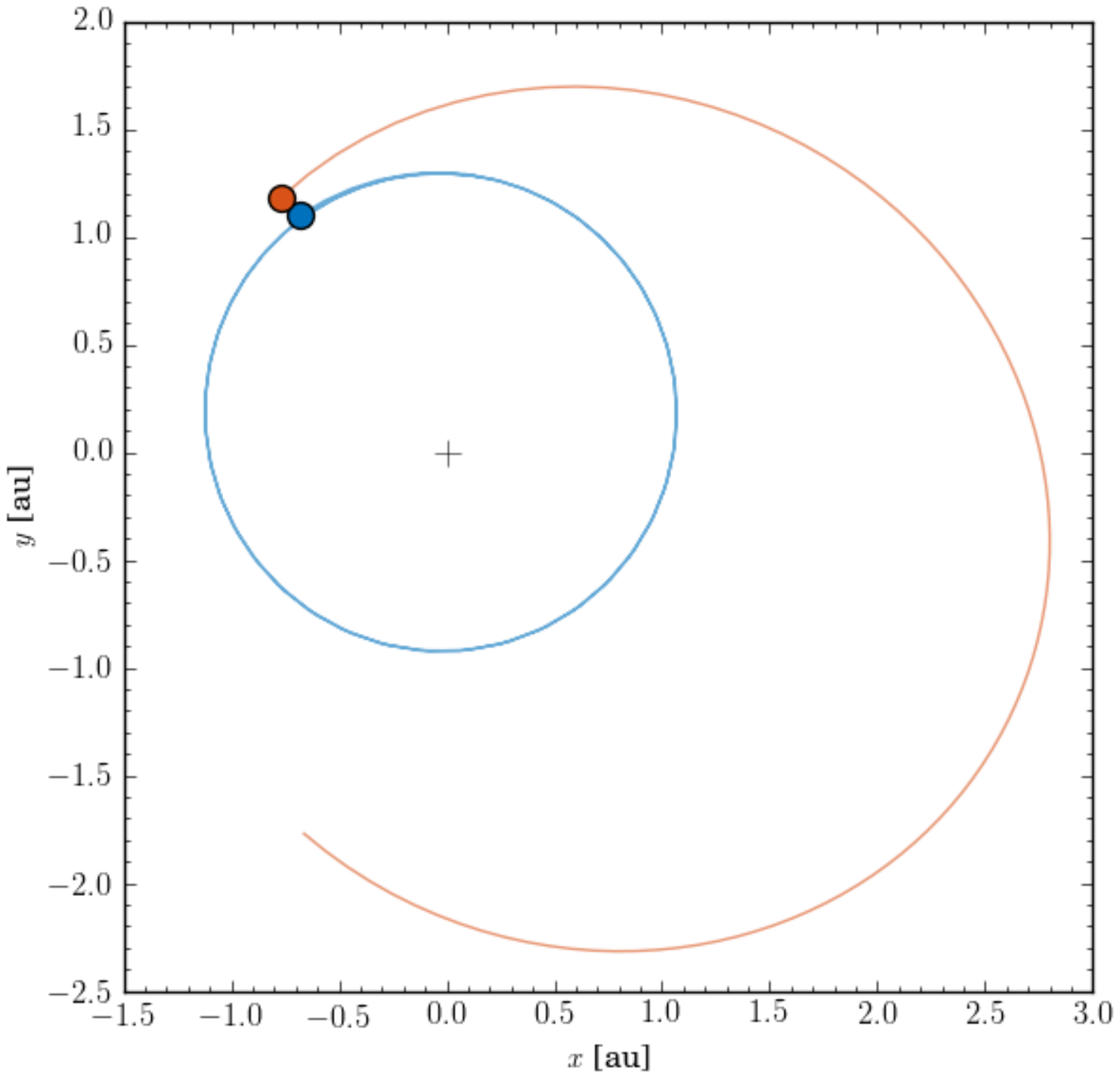}
    \includegraphics[width=0.33\textwidth,trim={0cm 5cm 1cm 5cm},clip]{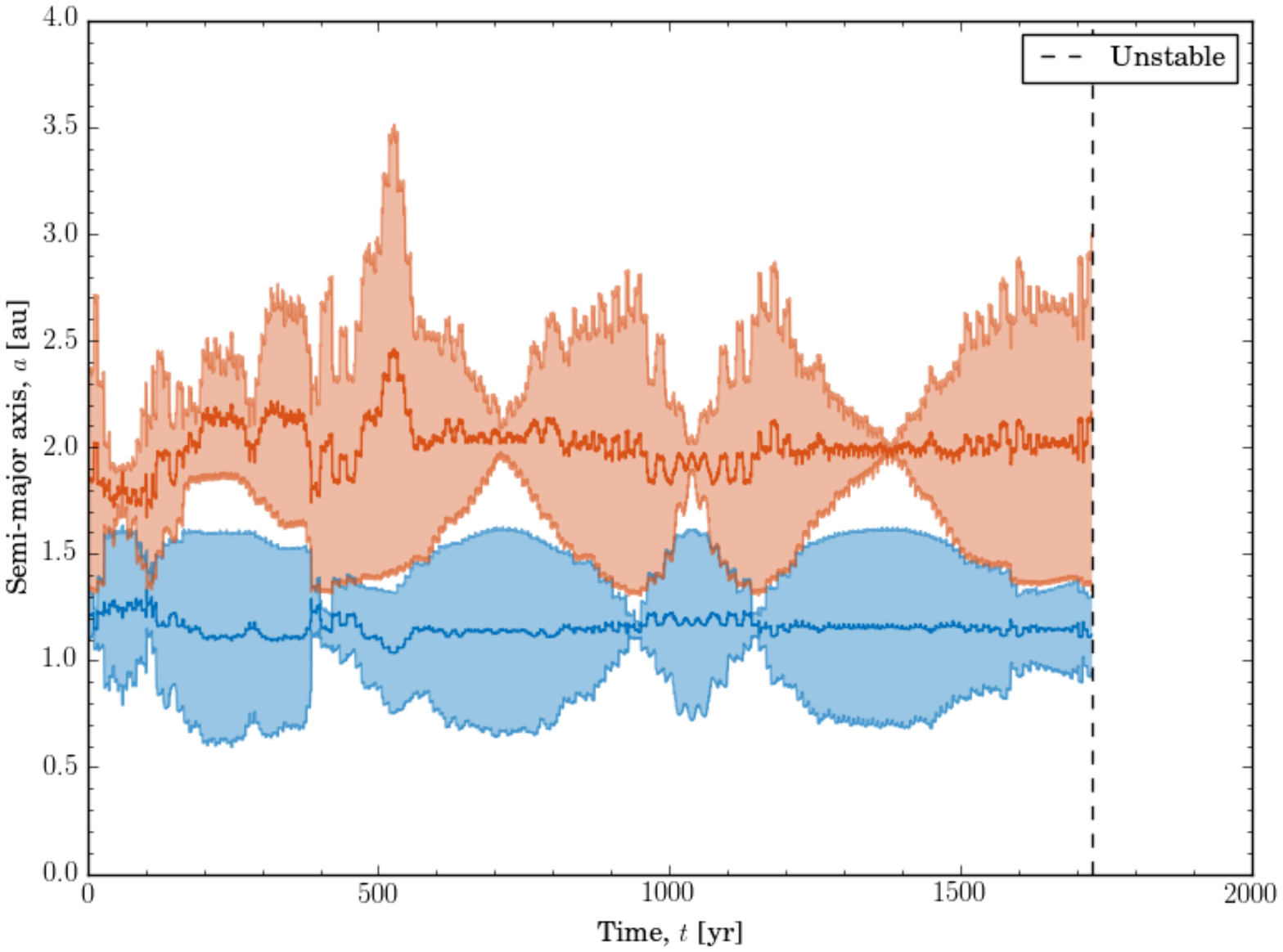}
    \includegraphics[width=0.34\textwidth,trim={0cm -2cm 0cm 0cm},clip]{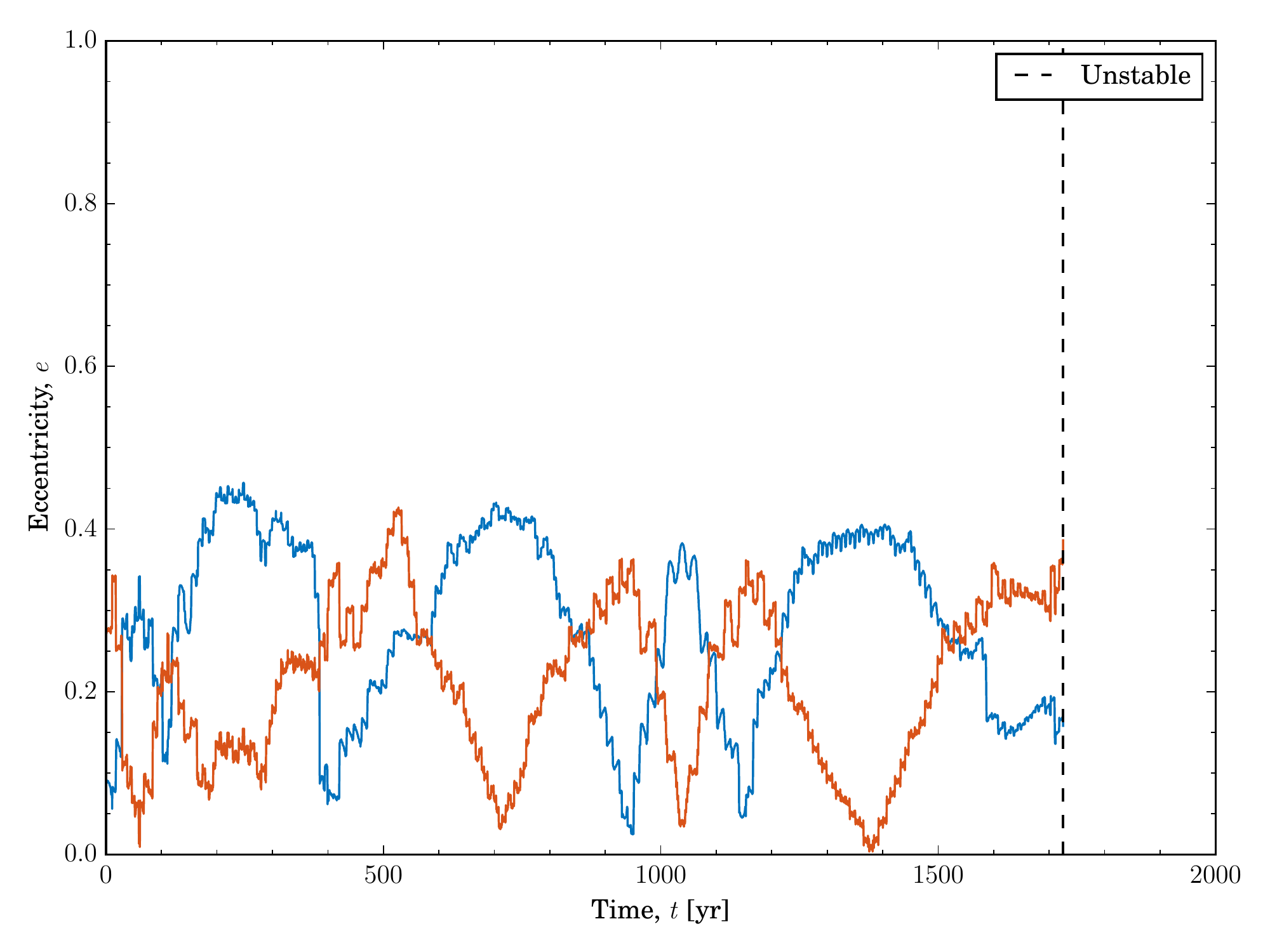}
    \caption{Here we illustrate the instability of the S16 orbital solution as demonstrated by our dynamical simulations. \textit{Left:} {A schematic plot of the orbital evolution of planets b and c showing the final complete orbit of innermost planet, HD~47366b, prior to the point at which instability occurs. Showing the entire evolution of the planets for the first ~1700 years is just a clump of overlapping, precessing orbits. This schematic just shows the lead up to the moment of instability.} In this plot the red dot and line denote the position of HD~47366c and its orbit, whilst the blue dot and line denote the position of HD~47366b and its orbit. \textit{Middle:}  The semi-major axis evolution prior to instability (vertical dashed line). The red line denotes the semi-major axis of HD~47366c, whilst the blue line denote the semi-major axis of HD~47366b. Shaded regions denote the apastron and periastron distances at each time interval for the two planets. \textit{Right:} {The eccentricity evolution of the two planets until instability occurs (vertical dashed line). The red line denotes the eccentricity of HD~47366c, whilst the blue line denotes the eccentricity of HD~47366b.} \label{fig:SystemEvolution}}
\end{figure*}

As a further illustration of the S16 orbital solution, the results of our dynamical simulations for best-fit architecture proposed in S16 are shown in Figure~\ref{fig:SystemEvolution}. In these plots we show the evolution of the semi-major axis and orbital eccentricity of the two planets as a function of time, along with a schematic plot of the proposed system architecture. It is quite self evident that the system as proposed in that work hits dynamical instability very quickly, after a period of under 10,000 years. 

Taken in concert, the results of our dynamical analysis suggest that 
the system as published in S16 is unlikely to be dynamically feasible. 
The observational data, however, do show two 
strong signals, and so it seems highly likely that the proposed planets 
really exist. Therefore, we revisit the fitting process for 
the system, to see whether the data could be fit equally well by any 
alternative solutions.

\subsection{Refitting the data}

Given that the HD~47366 planets appear to be dynamically interacting, a 
Keplerian fit as performed in S16 does not fully account for the system 
behaviour. For systems in full or near orbital resonances, 
mutual interactions are an important piece of physics to include \citep{1999Chambers,2006Laughlin}. 
We have carried out a comparative Levenberg-Marquardt and Bayesian 
data analysis on a binned data set. 
For both modelling techniques our analysis includes the full three-body gravitational 
interactions, from which we robustly derive a best fit solution for the architecture 
of the system.

We have adopted the stellar mass of 
2.19~$M_{\odot}$ derived in \cite{2016Wittenmyer}.  As S16 used 1.81~$M_{\odot}$, 
this simply changes the scale of the system but does not affect the 
overall dynamical behaviour.

Prior to modelling we binned the data on dates with multiple 
successive observations, adopting the weighted mean 
value of the velocities in each visit. The error bar of each binned 
point was calculated as the quadrature sum of the r.m.s. about the mean and 
the mean internal uncertainty.

\subsubsection{Levenberg-Marquardt approach}

In the first approach we use the Runge-Kutta integrator within {\sc Systemic} \citep{2009Meschiari}. {The {\sc Systemic} Console has the ability to account for interactions between planets to produce a self-consistent Newtonian fit. We use this 4$^{th}$/5$^{th}$ order Runge-Kutta approach with adaptive timestep control to model the planets in the HD~47366 system.  Given the degree to which the planets `talk' to each other, as evidenced by our initial dynamical simulations, we felt it prudent to adopt this fitting technique.} Since {\sc Systemic} can only fit a maximum of five data sets simultaneously, we merged  the CES-O and CES-N data by applying the 24.7~$ms^{-1}$\ relative velocity offset between them as obtained by S16.      

\begin{figure*}
    \centering
    \subfigure{\includegraphics[width=0.33\textwidth,trim={1cm 0cm 1cm 0cm},clip]{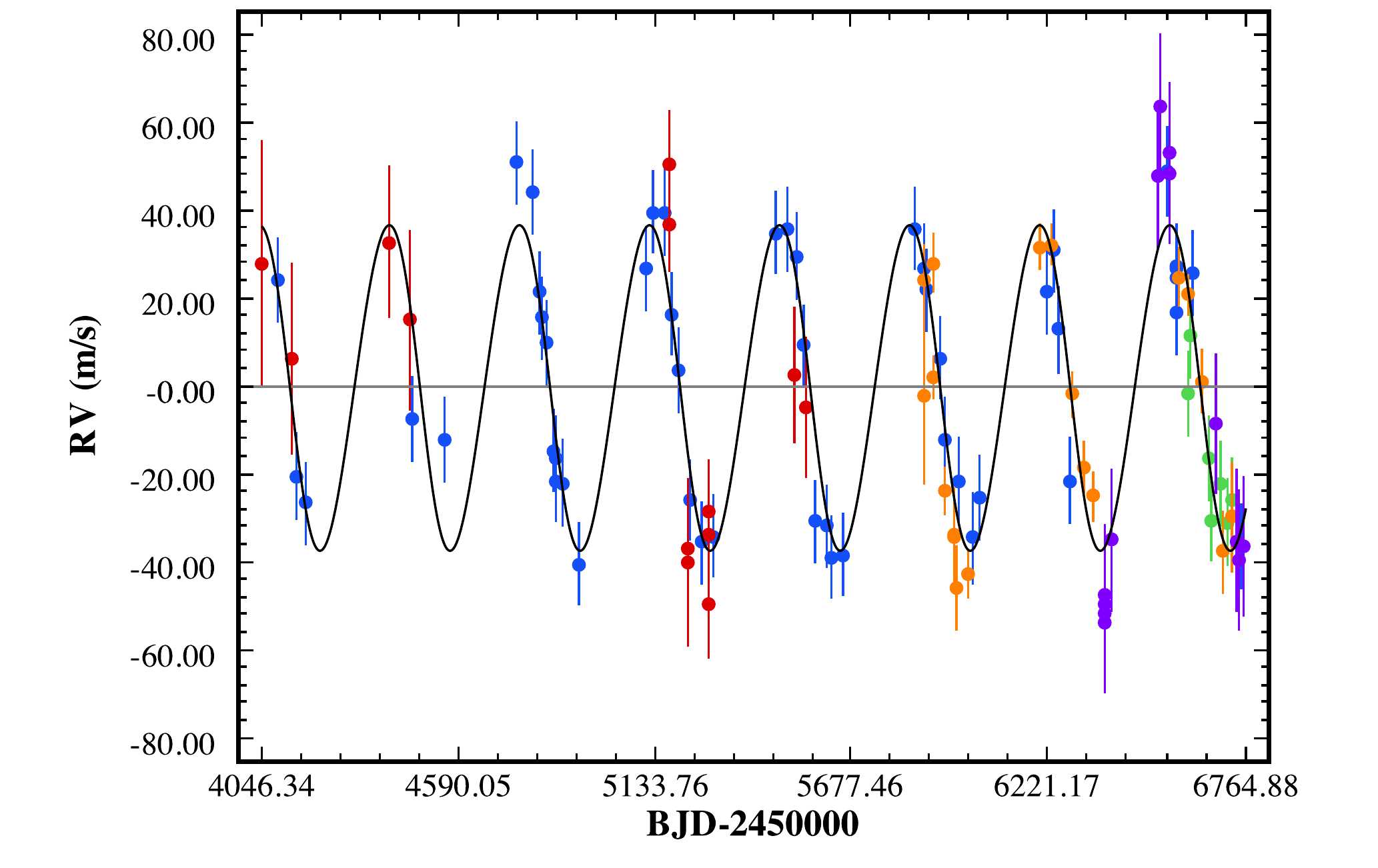}}
    \subfigure{\includegraphics[width=0.33\textwidth,trim={1cm 0cm 1cm 0cm},clip]{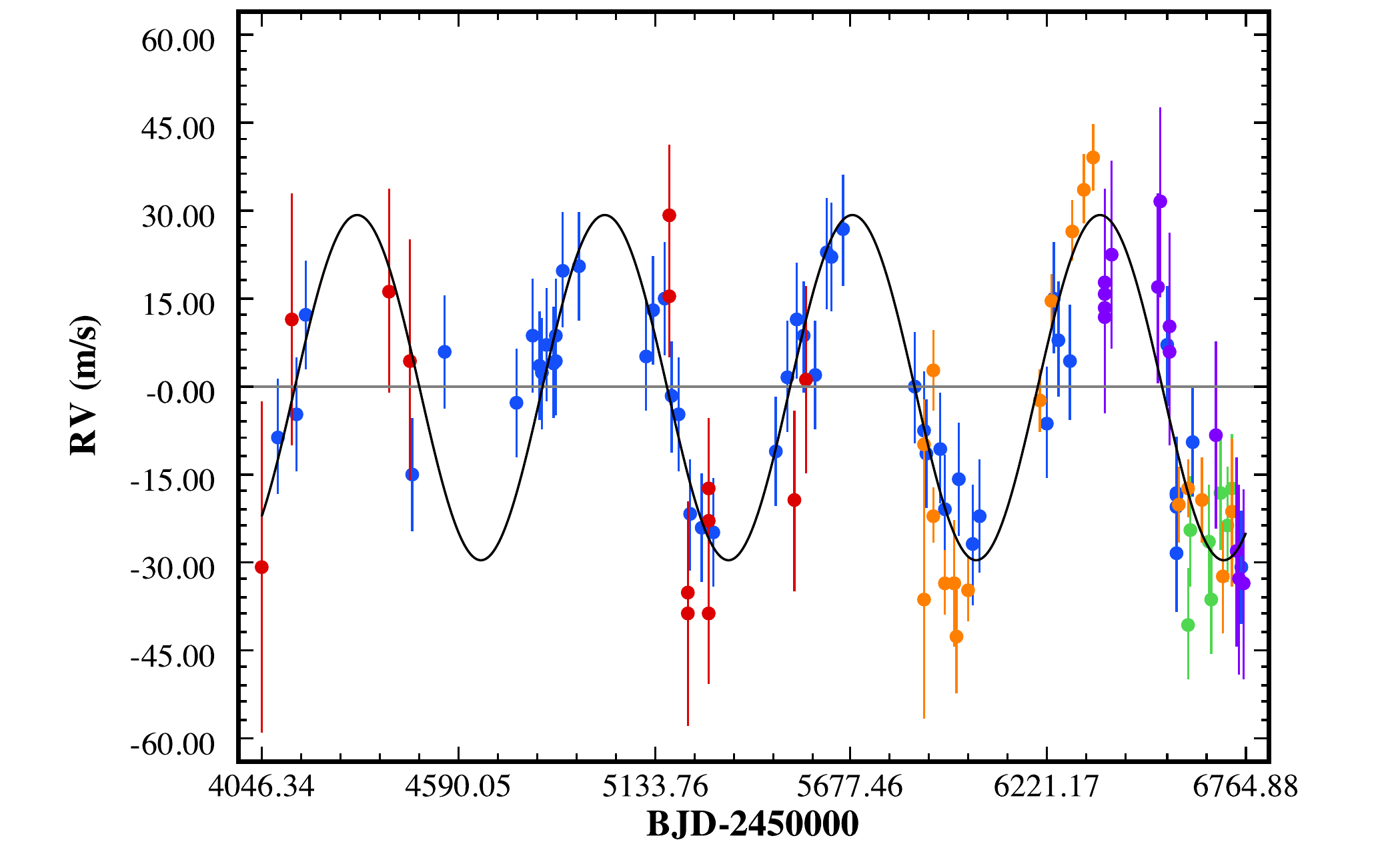}}
    \subfigure{\includegraphics[width=0.33\textwidth,trim={1cm 0cm 1cm 0cm},clip]{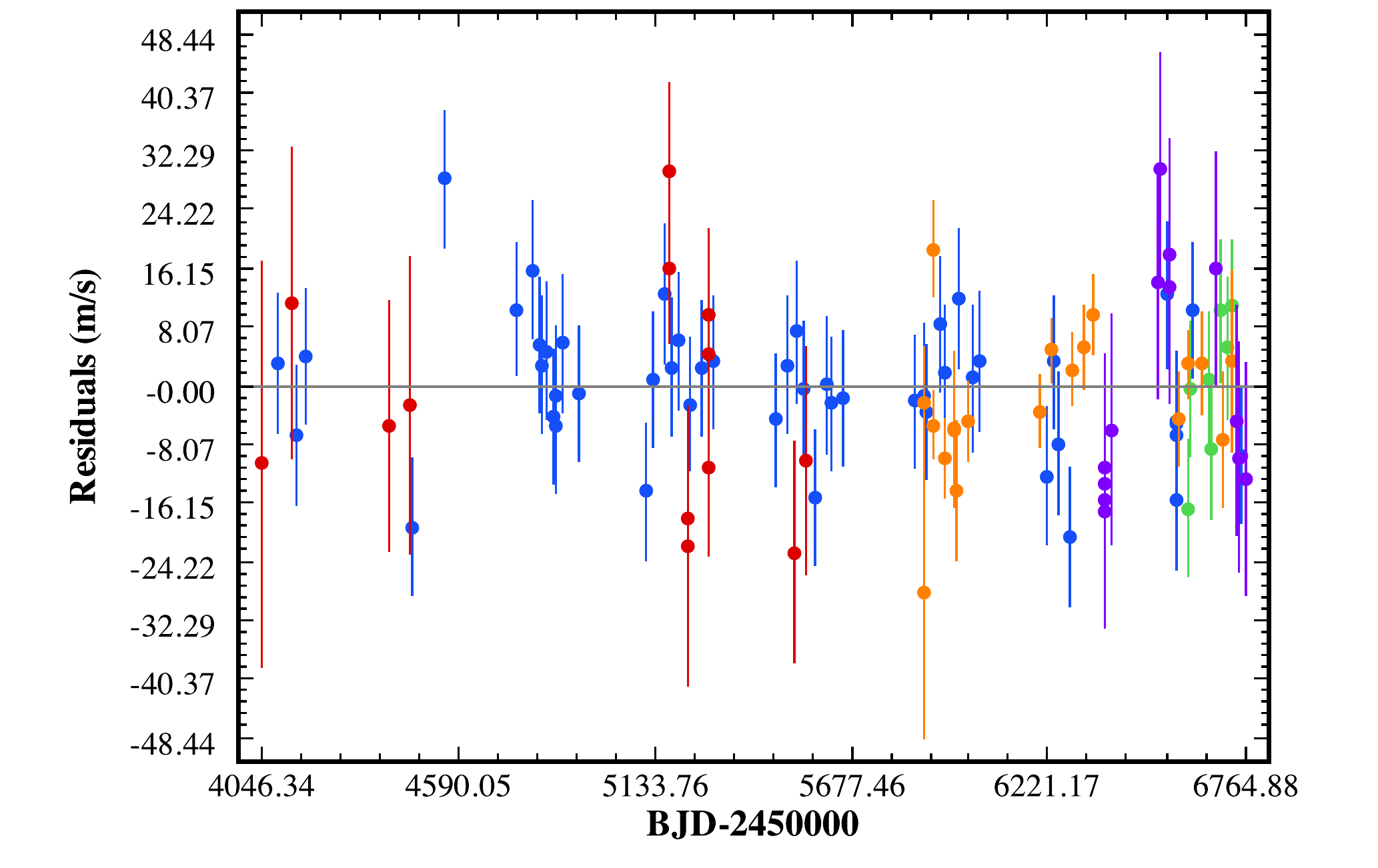}}
    \caption{Left: Data and model fit (black solid curve) for HD~47366b. Middle: Same, but for 
    the outer planet HD~47366c. Right: Residuals to the model fit. Data sets are represented by coloured points: Blue -- HIDES-S, red -- CES-N and CES-O, orange -- HRS, purple -- AAT, green -- HIDES-F. \label{fig:LMnewfitplots}}
\end{figure*}

The best fit results are given in Table~\ref{tab:LMfit} and the orbit fits are shown in 
Figure~\ref{fig:LMnewfitplots}.  Parameter uncertainties are obtained from a 
MCMC chain with $10^7$ steps, with the quoted 1-$\sigma$ uncertainties 
representing the range between the 15.87 and 84.13 percentiles of the 
posterior distribution.  The reduced $\chi^2$ is 1.35 and the residual 
r.m.s. about the fit is 11.4~$ms^{-1}$, as compared to the Keplerian fit 
obtained by S16 (r.m.s. = 14.7~$ms^{-1}$ and reduced $\chi^2~=~1.0$ by 
construction).  

\begin{deluxetable}{lcc}
\tablewidth{0.4\textwidth}
\tablecaption{Orbital parameters for the HD~47366 system based on Levenberg-Marquardt analysis. The time of periastron passage is a fit parameter; mean anomaly, mass $m\sin i$, and semi-major axis $a$ are derived parameters. \label{tab:LMfit}}
\tablecolumns{3}
\tablehead{
\colhead{Parameter}  & \colhead{HD~47366 b}  & \colhead{HD~47366 c} \\
}
\startdata
$P$ (d)    & 360.2$^{+3.4}_{-3.9}$  &  686.4$^{+13.7}_{-8.1}$  \\
$T_{\rm peri}$ (BJD-2450000) & 3796~$\pm$~51 & 3817~$\pm$~57 \\
$K$ (m/s) & 34.5$^{+3.1}_{-2.6}$  & 29.7$^{+2.1}_{-1.8}$  \\
$e$        & 0.10$^{+0.05}_{-0.05}$  & 0.18$^{+0.06}_{-0.06}$  \\
$\omega$ (\degr) & 134$^{+47}_{-46}$  & 107$^{+24}_{-24}$  \\
\hline
Mean anomaly, $M$ (\degr)  & 250$^{+41}_{-51}$  & 120$^{+30}_{-28}$ \\
$m \sin i$ ($m_{\rm Jup}$) & 2.03$^{+0.18}_{-0.15}$  & 2.14$^{+0.15}_{-0.13}$ \\
$a$ (au)   & 1.287$^{+0.008}_{-0.010}$  & 1.978$^{+0.026}_{-0.016}$  \\ 
\enddata
\raggedright
\end{deluxetable}

\begin{figure*}
    \centering
    \subfigure{\includegraphics[width=0.18\textwidth,trim={1cm 0.5cm 0.5cm 0.5cm },clip]{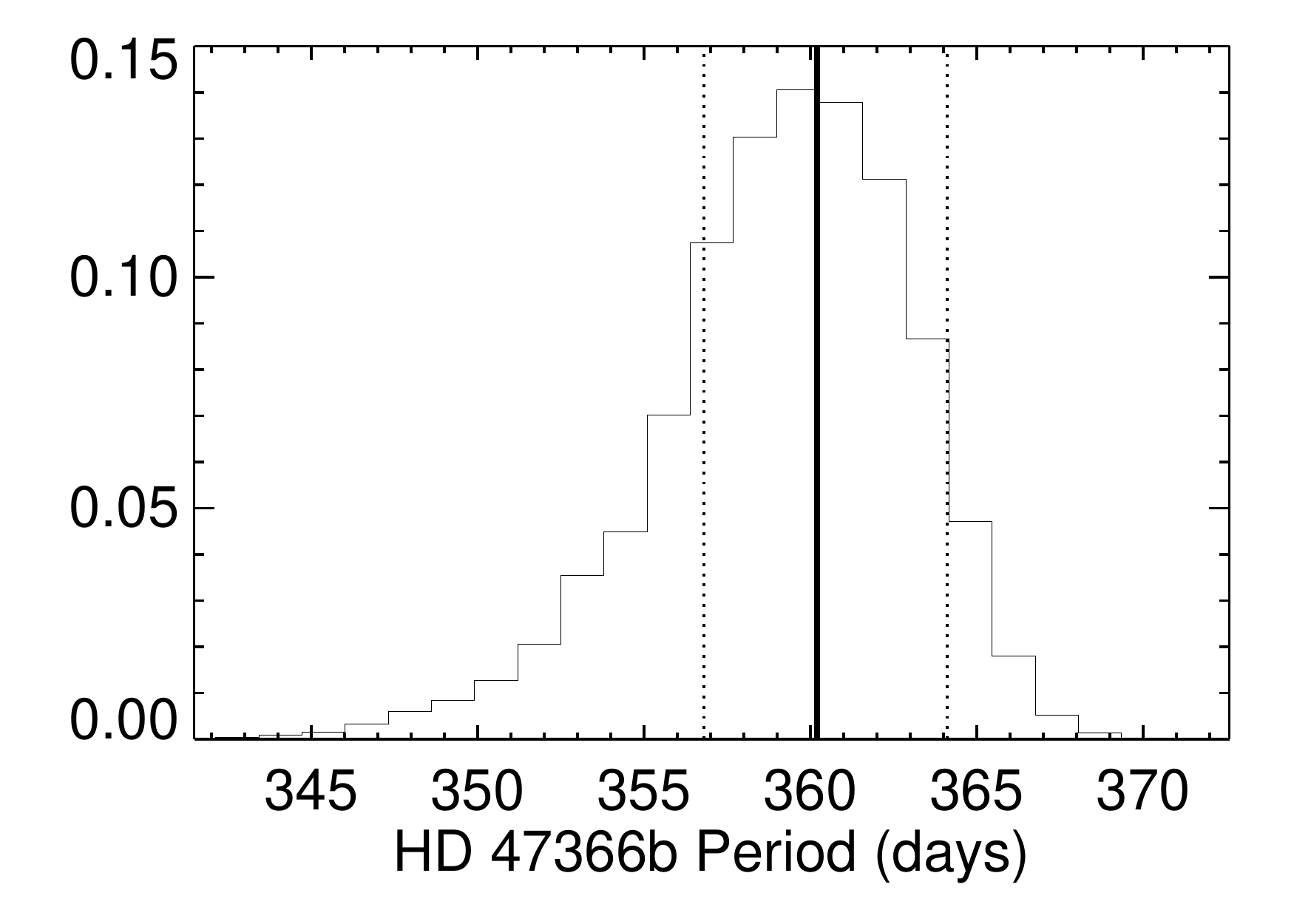}}
    \subfigure{\includegraphics[width=0.18\textwidth,trim={1cm 0.5cm 0.5cm 0.5cm },clip]{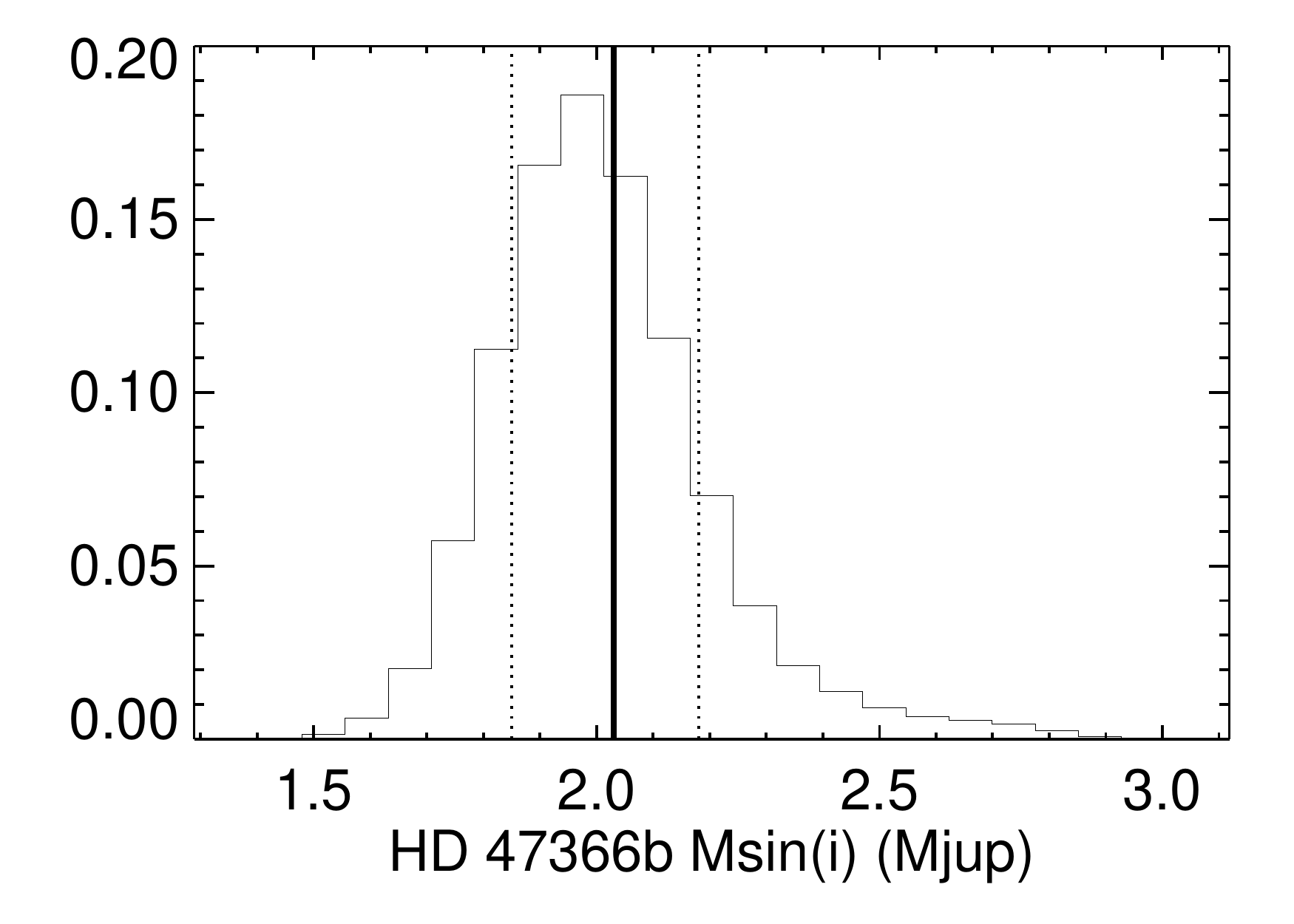}}
    \subfigure{\includegraphics[width=0.18\textwidth,trim={1cm 0.5cm 0.5cm 0.5cm },clip]{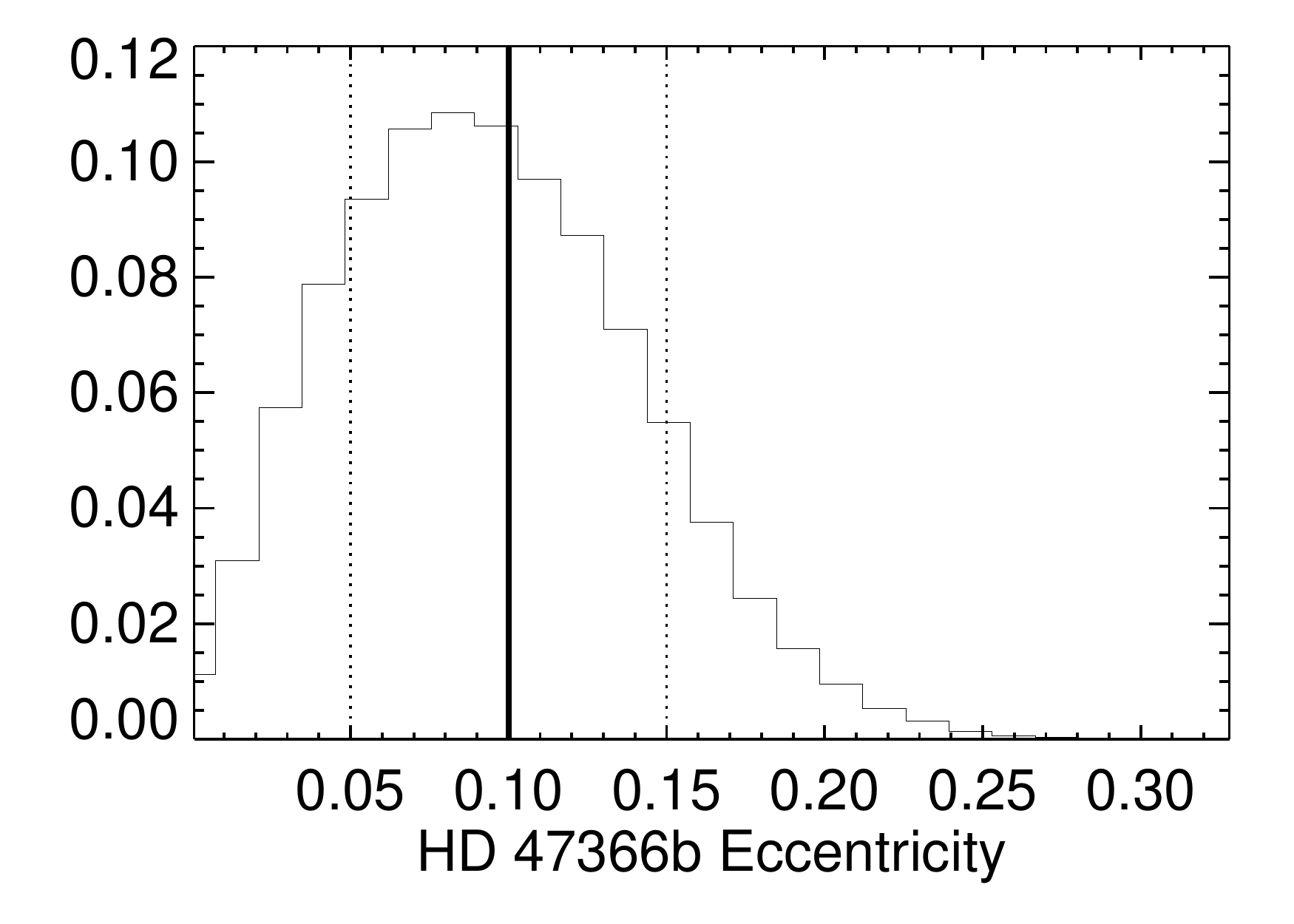}}
    \subfigure{\includegraphics[width=0.18\textwidth,trim={1cm 0.5cm 0.5cm 0.5cm },clip]{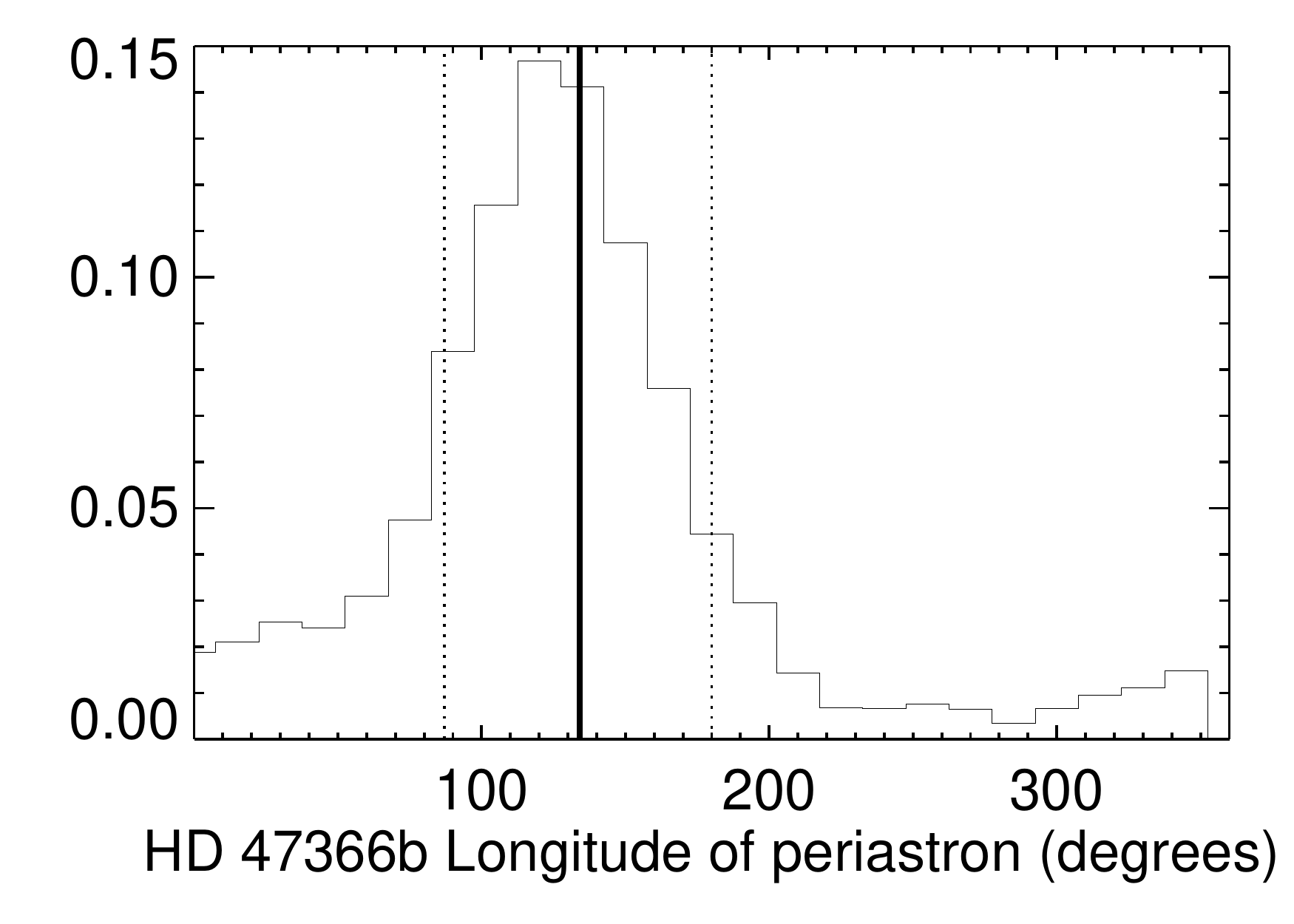}}
    \subfigure{\includegraphics[width=0.18\textwidth,trim={1cm 0.5cm 0.5cm 0.5cm },clip]{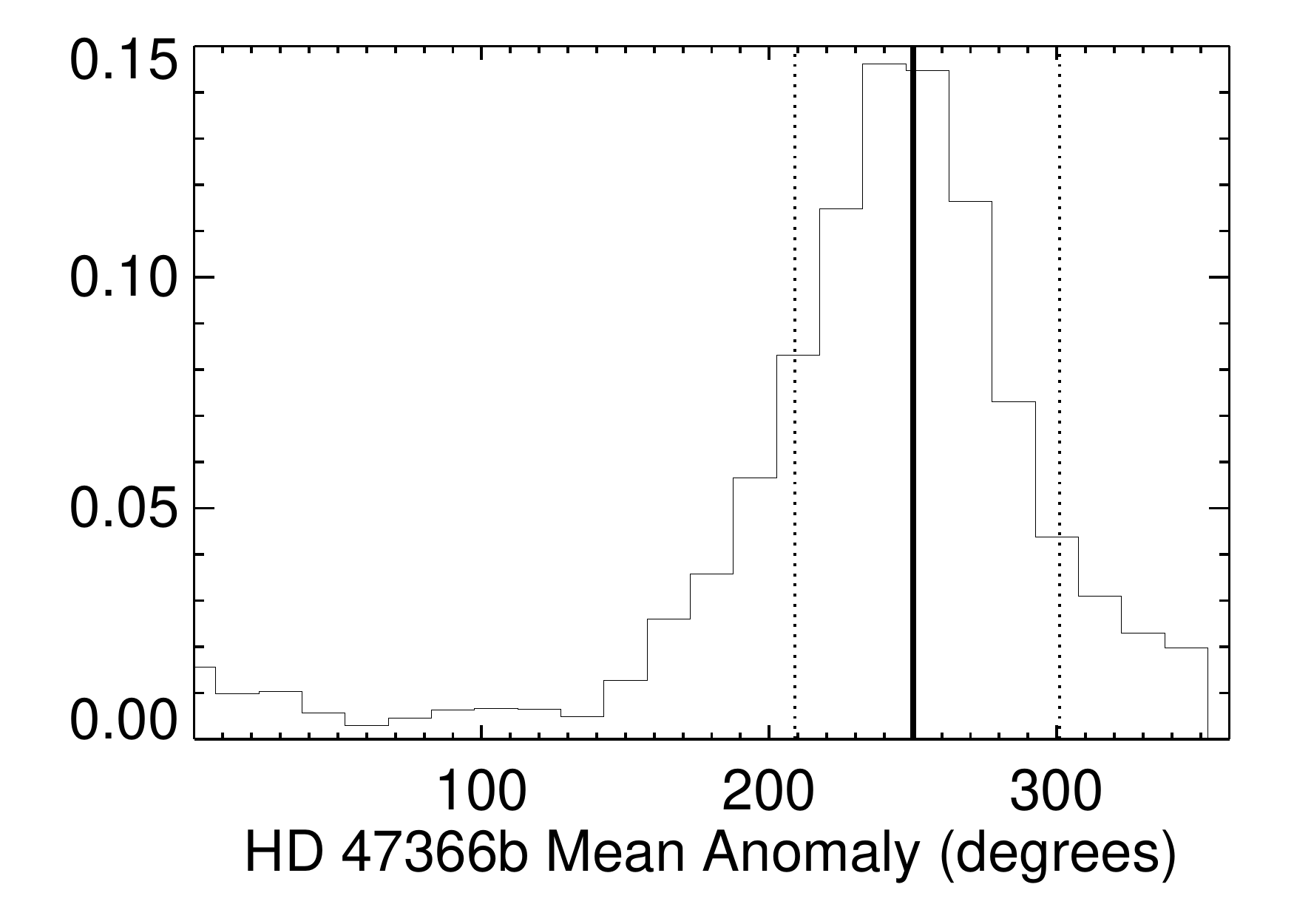}}\\
    \subfigure{\includegraphics[width=0.18\textwidth,trim={1cm 0.5cm 0.5cm 0.5cm },clip]{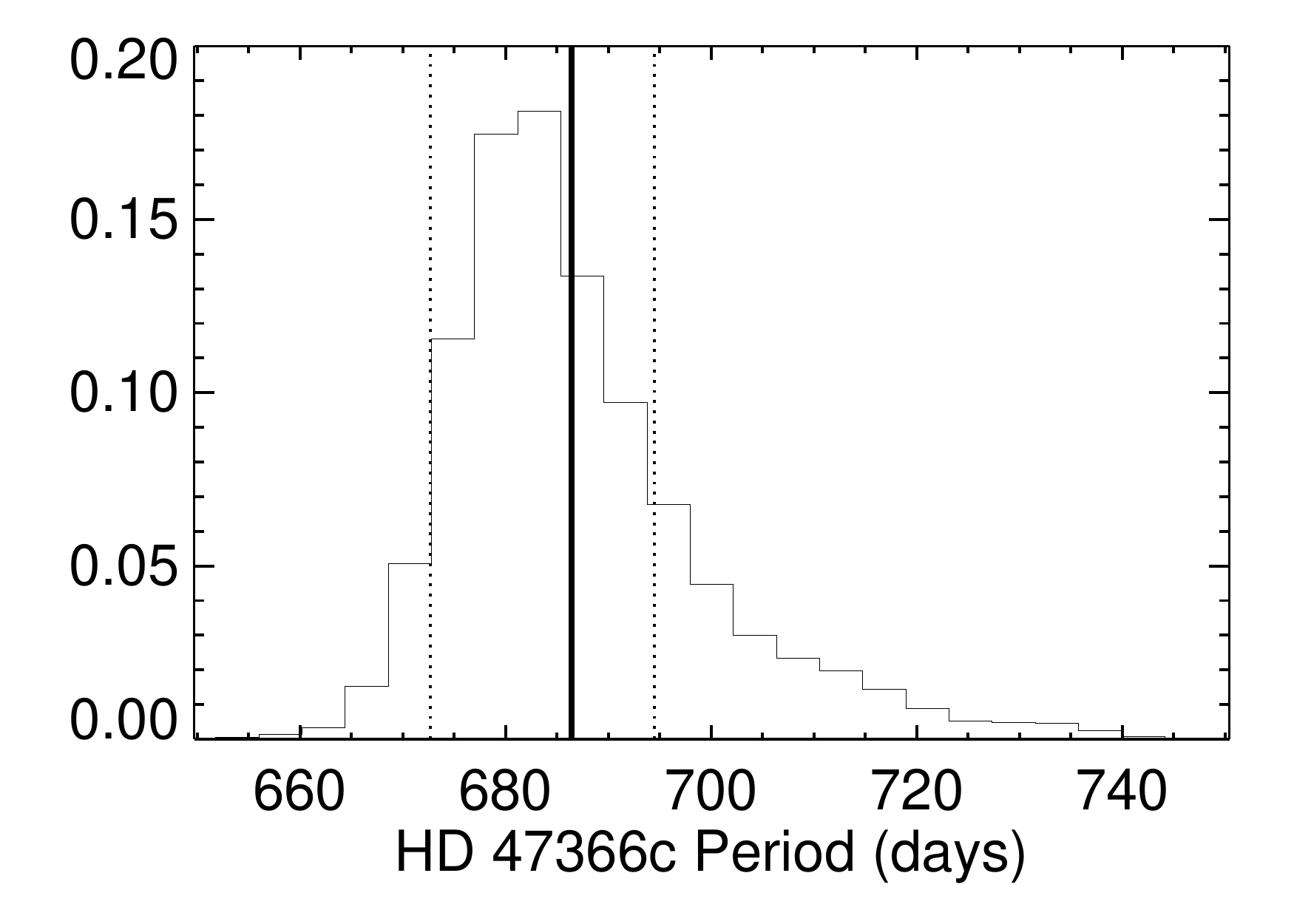}}
    \subfigure{\includegraphics[width=0.18\textwidth,trim={1cm 0.5cm 0.5cm 0.5cm },clip]{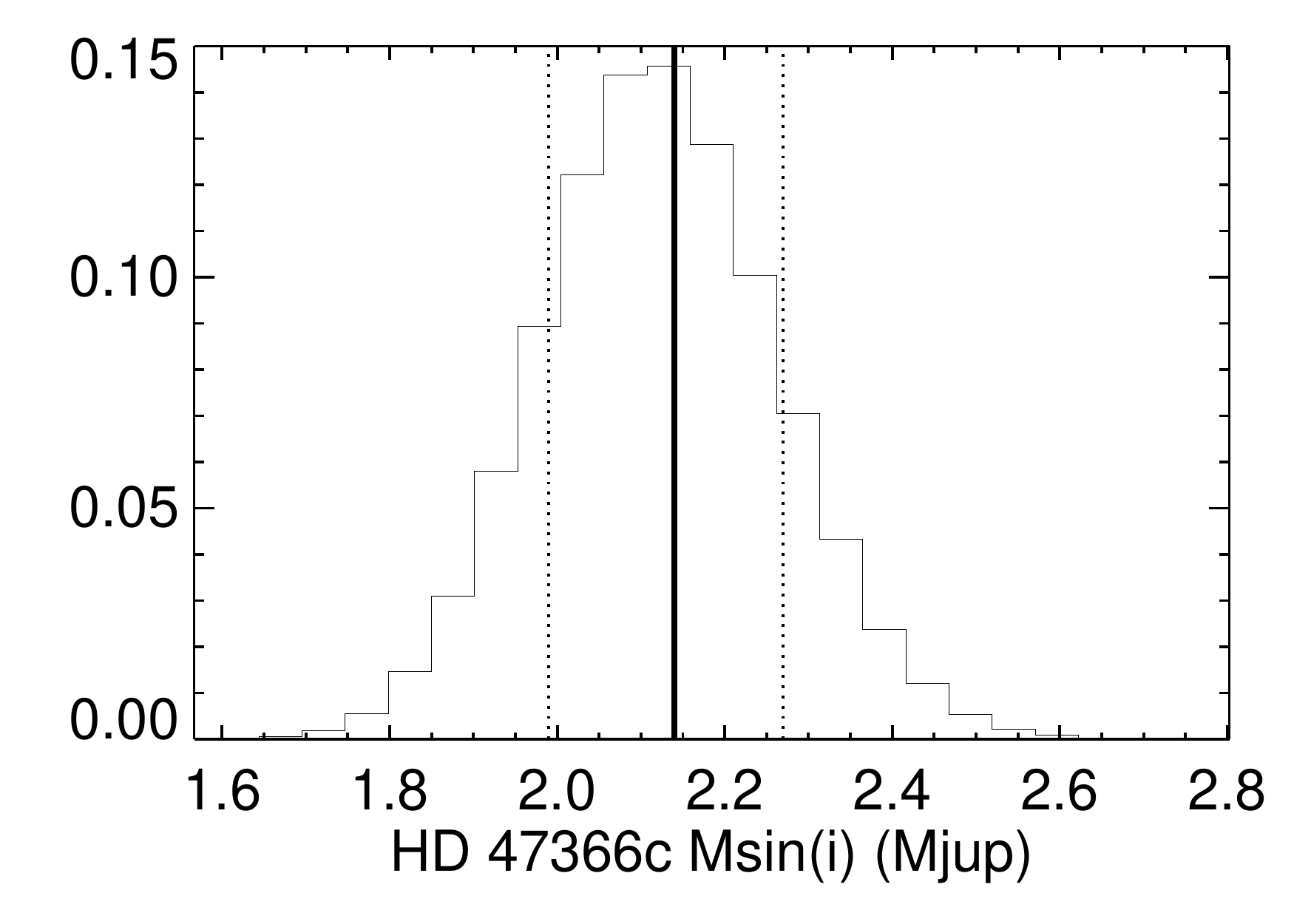}}
    \subfigure{\includegraphics[width=0.18\textwidth,trim={1cm 0.5cm 0.5cm 0.5cm },clip]{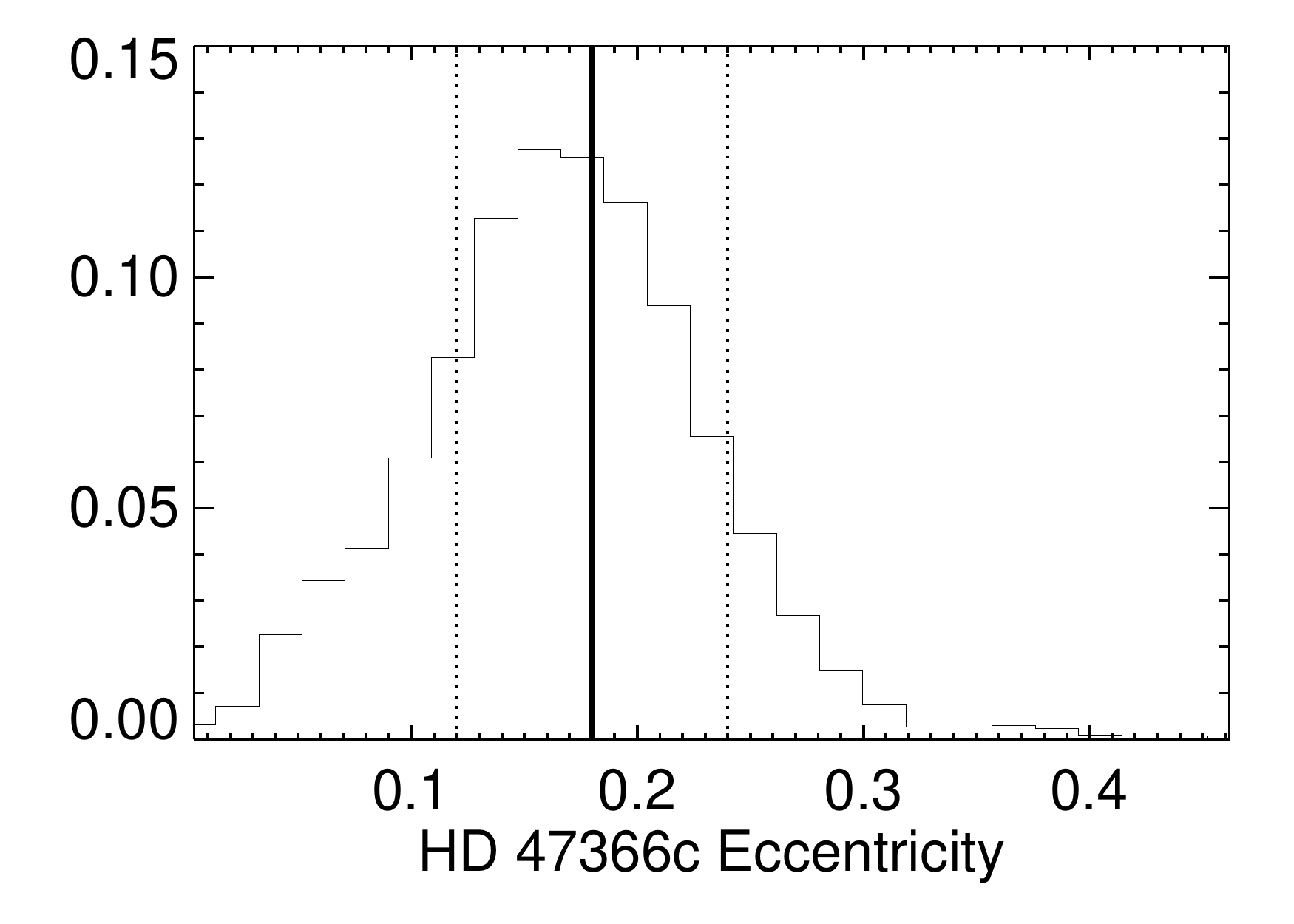}}
    \subfigure{\includegraphics[width=0.18\textwidth,trim={1cm 0.5cm 0.5cm 0.5cm },clip]{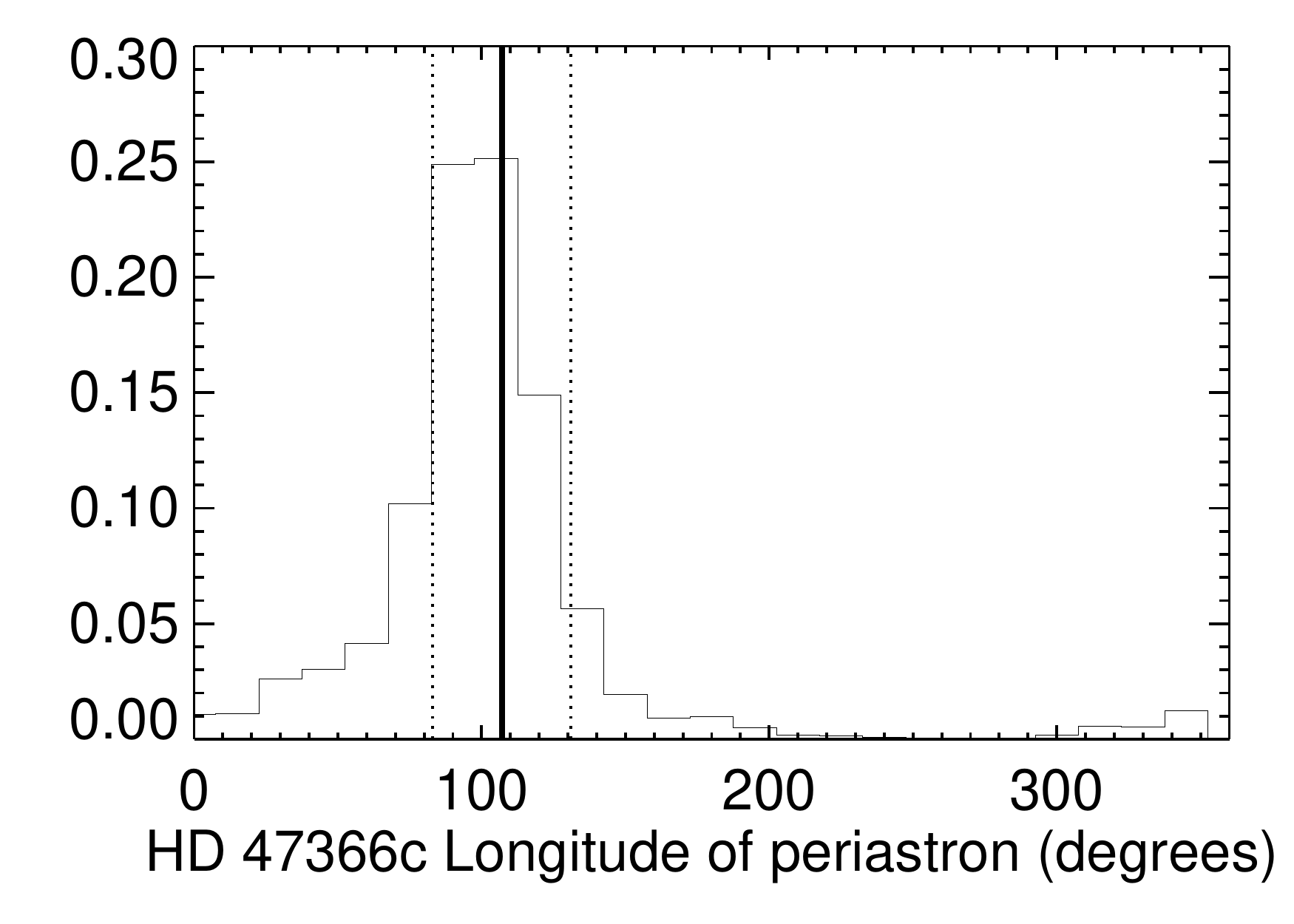}}
    \subfigure{\includegraphics[width=0.18\textwidth,trim={1cm 0.5cm 0.5cm 0.5cm },clip]{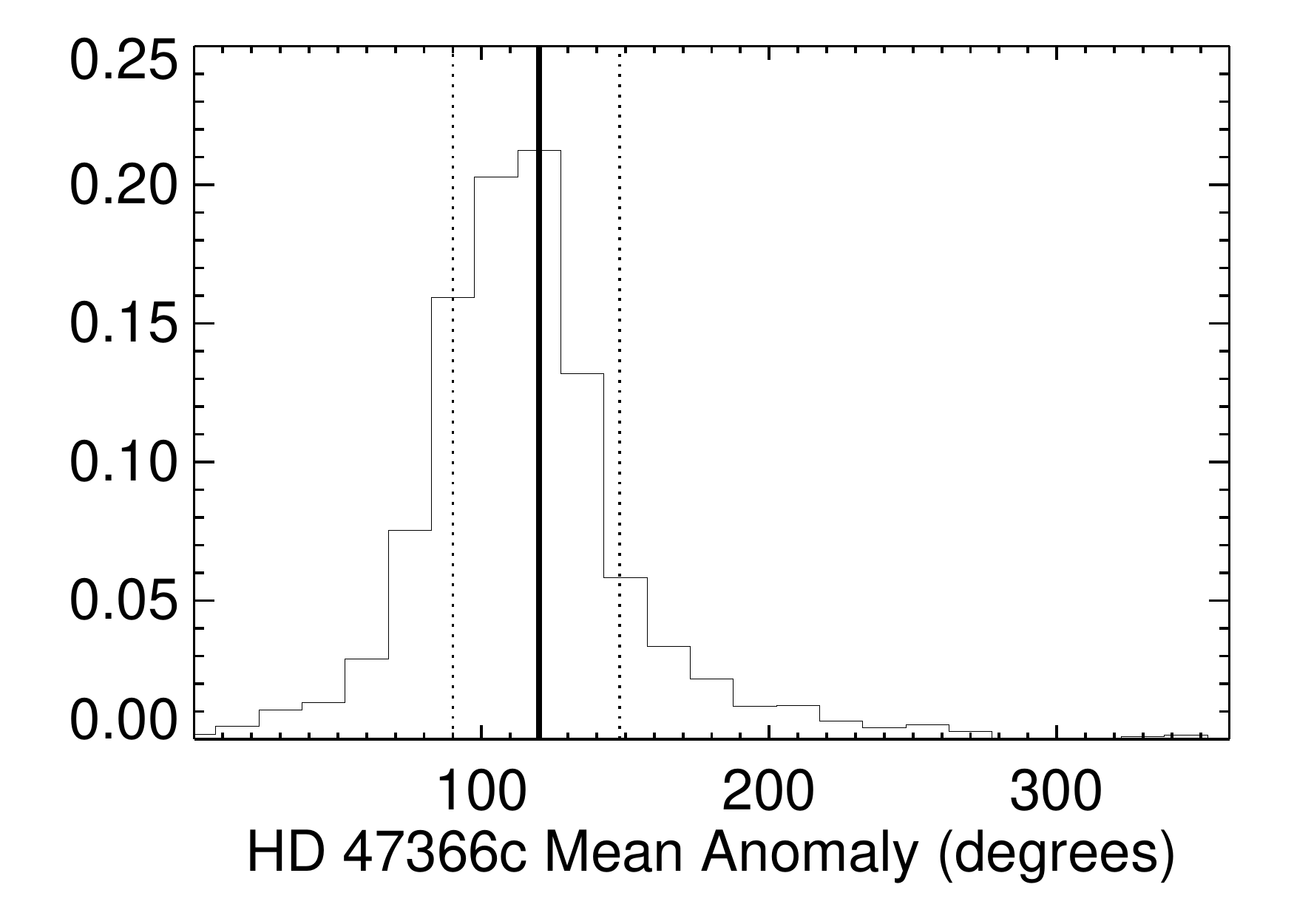}}\\
    \caption{Posterior probability distributions for HD~47366b (top) and HD~47366c (bottom) from our refit of orbital parameters using the combined Levenberg-Marquardt and MCMC analysis. Parameters are (in order left to right) orbital period ($P$), line-of-sight mass ($m\sin i$), eccentricity ($e$), longitude of periastron ($\omega$), and mean anomaly ($M$). The distributions are well behaved, and the simultaneous best-fit values for the parameters of both planets lie close to the peak of their respective probability distributions. \label{fig:LM_posterior}}
\end{figure*}

The posterior distributions of parameters for the two planets are shown 
in Figure~\ref{fig:LM_posterior}. Modelling the binned data, we find the overall 
fit has improved in r.m.s. scatter compared to 
that of S16, and the posterior probability distributions are now 
unimodal.  Our re-fit also results in a lower eccentricity for the outer 
planet; a critical criterion for dynamical stability (and hence 
viability) of the HD~47366 system.

\subsubsection{Bayesian approach}

In our second effort, we obtained posterior distributions of the HD~47366 system's orbital parameters ultising the Markov Chain Monte Carlo (MCMC) Bayesian code Exoplanet Mcmc Parallel tEmpering Radial velOcity fitteR\footnote{\href{https://github.com/ReddTea/astroEMPEROR}{https://github.com/ReddTea/astroEMPEROR}} ({\sc astroemperor}, Jenkins \& Pena, in prep.). As described in \citep{Giant3}, {\sc astroemperor} utilises thermodynamic integration methods \citep{2005Gregory} following an affine invariant MCMC engine, performed using the {\sc Python} {\sc emcee} package \citep{2013ForemanMackey}. {Using an affine invariant algorithm such as {\sc emcee} allows the MCMC analysis to perform equally well under all linear transformations consequently being insensitive to covariances among the fitting parameters \citep{2013ForemanMackey}.} First-order moving average models are used within {\sc astroemperor} to measure the correlated noise within the radial velocity measurements. A model selection is performed automatically by {\sc emperor}, whereby an arbitrary Bayes Factor value of five is required. {This means a threshold probability of 150 is needed for a more complex model to be favored over a less complex one.} The {\sc astroemperor} code also automatically determines which of the orbital parameters, such as period and amplitude, are statistically significantly different from zero, {with the Bayesian information criterion (BIC), Akaike information criterion (AIC) and maximum a posteriori probability (MAP) estimate values} calculated for each planetary signal. Flat priors are applied to all parameters except for the eccentricity and jitter priors that are folded Gaussian and Jeffries, respectively.

\begin{figure*}
    \centering
    \includegraphics[width=\textwidth]{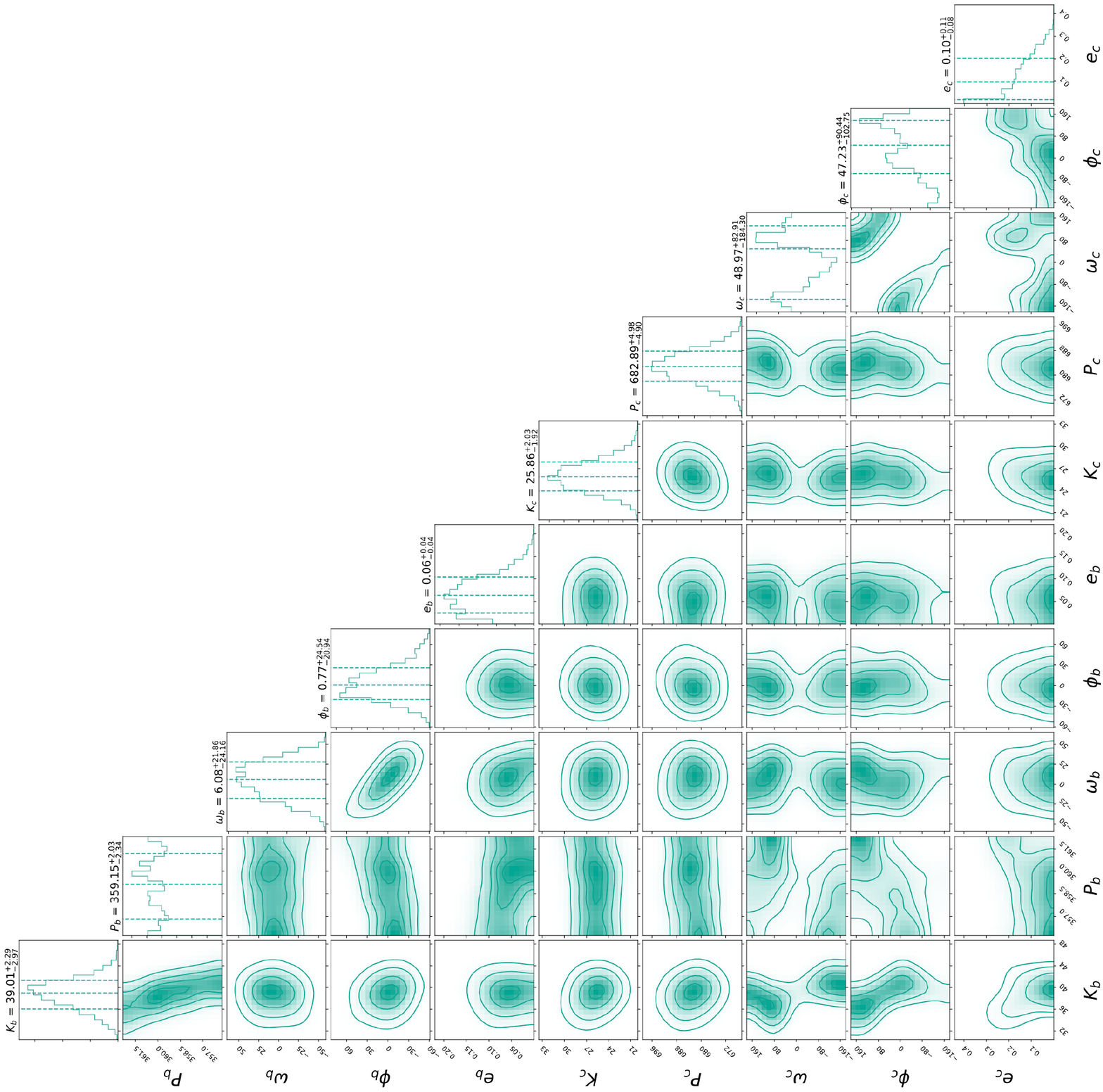}
    \caption{Bayesian posterior distributions of HD~47366 b and HD~47366 c's orbital parameters derived from {\sc astroemperor}. From left to right (top to bottom), the parameters are $K_{\rm b}$, $P_{\rm b}$, $\omega_{\rm b}$, $\phi_{\rm b}$, $e_{\rm b}$, $K_{\rm c}$, $P_{\rm c}$, $\omega_{\rm c}$, $\phi_{\rm c}$ and $e_{\rm c}$. Credible intervals are denoted by the solid contours with increments of 1-$\sigma$. Each 1-D histogram exhibits dashed lines, displaying the median and  $\pm \sigma$ values (also displayed above for clarity).}
    \label{fig:corner_plot}
\end{figure*}

For our particular analysis, the 'burn-in' chains were 3.75 million iterations long (5 temperatures, 150 walkers and 5,000 steps) with another 7.5 million chains exploring the parameter space thereafter (10,000 steps instead of 5,000). {\sc astroemperor} was implemented in an unbounded manner, from zero to two planetary signals, giving flexibility for the program to discover global minima within the parameter space. The results of the {\sc astroemperor} analysis are illustrated in Fig. \ref{fig:corner_plot} by a corner plot, summarised in Table \ref{tab:Bayesfit} and the BIC for each fit given in Table \ref{tab:Bayesstat}. Table \ref{tab:Bayesfit}'s results are based upon the posterior distribution's median value and the quoted 1-$\sigma$ values representing the 15.87 and 84.13 percentiles. The 1-D histograms of the Bayesian parameter fitting shows the fits are generally well-behaved and are relatively mono-modal. There is general good agreement between the values for the orbital parameters of the planets determined through both the Levenberg-Marquardt and Bayesian analyses of the data.

\begin{deluxetable}{lcc}
\tablewidth{0.4\textwidth}
\tablecaption{Bayesian re-fitting of the HD~47366 planetary system through the \textsc{Python} package \textsc{astroemperor}. \label{tab:Bayesfit}}
\tablecolumns{3}
\tablehead{
\colhead{Parameter}  & \colhead{HD~47366 b}  & \colhead{HD~47366 c} \\
}
\startdata
$P$ (d)          & 359.15~$^{+2.03}_{-2.34}$~  & 682.85~$^{+4.98}_{-4.90}$~ \\
$\phi$\footnote{$\phi$ is a measured parameter defined in \textsc{Astroemperor} as $\phi = M - \frac{2\pi}{P}t$, related to the mean anomaly (M), orbital period (P) and epoch time (t).} (\degr)   & 0.77~$^{+24.54}_{-20.94}$~  & 47.23~$^{+90.44}_{-102.75}$~ \\
$K$ (m/s)        & 39.01~$^{+2.29}_{-2.97}$~   & 25.86~$^{+2.03}_{-1.92}$~  \\
$e$              & 0.06~$^{+0.04}_{-0.04}$~    & 0.10~$^{+0.11}_{-0.08}$~ \\
$\omega$ (\degr) & 6.08~$^{+21.86}_{-24.16}$~  & 48.97~$^{+82.91}_{-184.30}$~ \\

\hline
$m \sin i$ ($m_{\rm Jup}$) & 2.30~$^{+0.13}_{-0.18}$~  & 1.88~$^{+0.12}_{-0.14}$~ \\
$a$ (au)   & 1.28~$^{+0.05}_{-0.06}$~  & 1.97~$^{+0.08}_{-0.09}$~ \\ 
\enddata
\raggedright
\end{deluxetable}

\begin{deluxetable}{lcc}
\tablewidth{0.4\textwidth}
\tablecaption{Bayesian information criterion (BIC) statistic for each {\sc astroemperor} signal fit from a zero ($k_0$) to two planet fit ($k_2$) \label{tab:Bayesstat}}
\tablecolumns{3}
\tablehead{
\colhead{Signals}  & \colhead{BIC}  & \colhead{$\Delta$ BIC~(k,k-1)}\\
}
\startdata
$k_0$ & 1142.71 & ...  \\
$k_1$ & 927.06 & 215.65  \\
$k_2$ & 817.80 & 109.26  \\
\enddata
\raggedright
\end{deluxetable}

\subsection{Dynamical simulations of the new solution}

With a new solution model available for the HD~47366 system, we repeated our 
earlier dynamical analysis. Our contextual runs again featured a 
hypercubic grid of 126,075 initial conditions in $a$--$e$--$\omega$--$M$ space, and 
our planet-pair cloud simulations again tested an additional 126,075 
solutions, centred on our newly found local minimum in $\chi^2$-space.

\begin{figure}
\includegraphics[width=0.35\textwidth,angle=270,trim={3cm 4cm 3cm 4cm},clip]{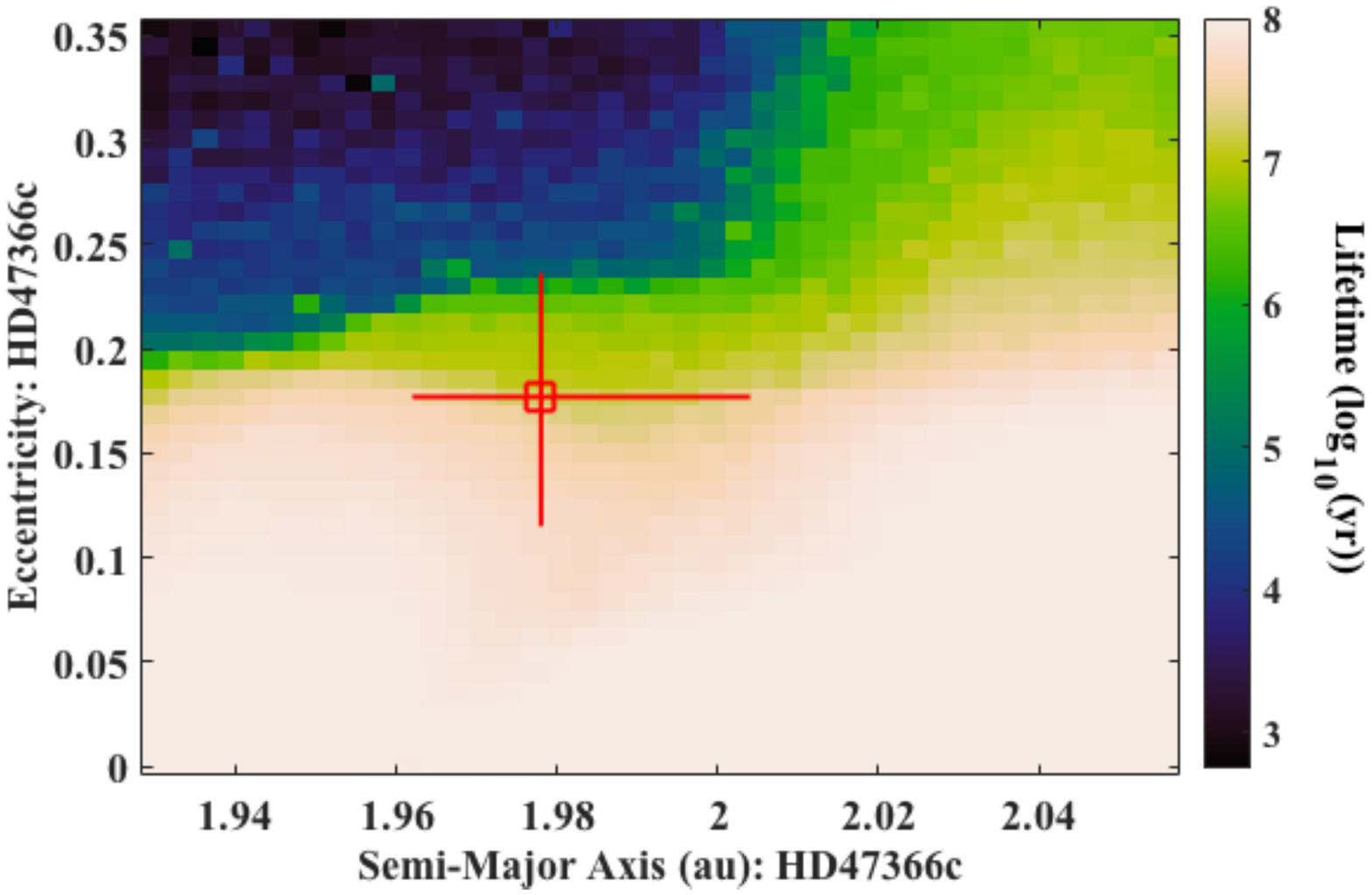}
\caption{The dynamical stability of the new solution for the orbits of 
the two planets around HD~47366, as a function of the initial orbit of 
HD~47366c.  The best-fit solution is again marked by the hollow red box, 
with the 1-$\sigma$ uncertainties denoted by the lines that radiate from 
within. As before, the lifetime shown at each location is the mean of 75 
trials. As a result of the reduced orbital eccentricity in the new fit, 
the solution now lies at the edge of the broad stable region, with many 
trials within 1-$\sigma$ of the best-fit surviving for the full 100~Myr 
duration of our integrations.
\label{fig:RobContext}}
\end{figure} 

The results of our contextual simulations for this new solution can be 
seen in Figure~\ref{fig:RobContext}. As a result of the new, reduced orbital 
eccentricity for this solution, the best-fit to the data now lies on the 
edge of a broad region of dynamical stability. A significant fraction of 
the individual trials within the 1-$\sigma$ uncertainty range on the 
solution were found to survive for the full 100~Myr duration of our 
integrations, a result in stark contrast with those we performed of the 
S16 solution. We note, in passing, that whilst the general 
structures visible in Figure~\ref{fig:RobContext} are the same as those in 
Figure~\ref{fig:SatoContext}, the broad expanse of solutions within the 2:1 
mean-motion resonance between the two planets (to the right of the plot) 
now exhibit somewhat improved stability. This is the result of the 
broader range of $\omega$ and $M$ values sampled by the new solution, 
which increases the likelihood of the two planets being trapped in a 
stable resonant configuration in our runs. 

\begin{figure*}
    \includegraphics[width=0.5\textwidth]{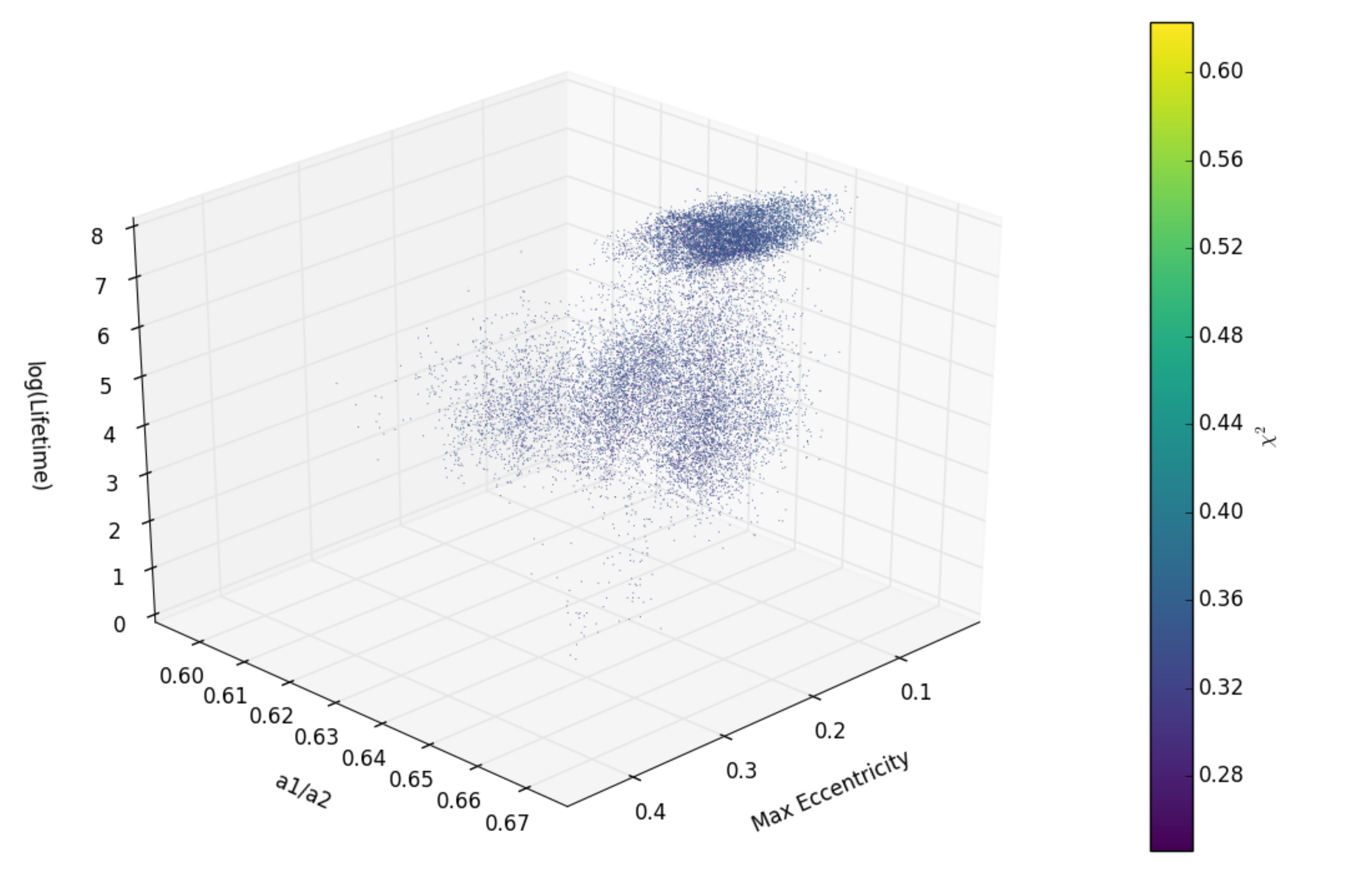}
	\includegraphics[width=0.5\textwidth]{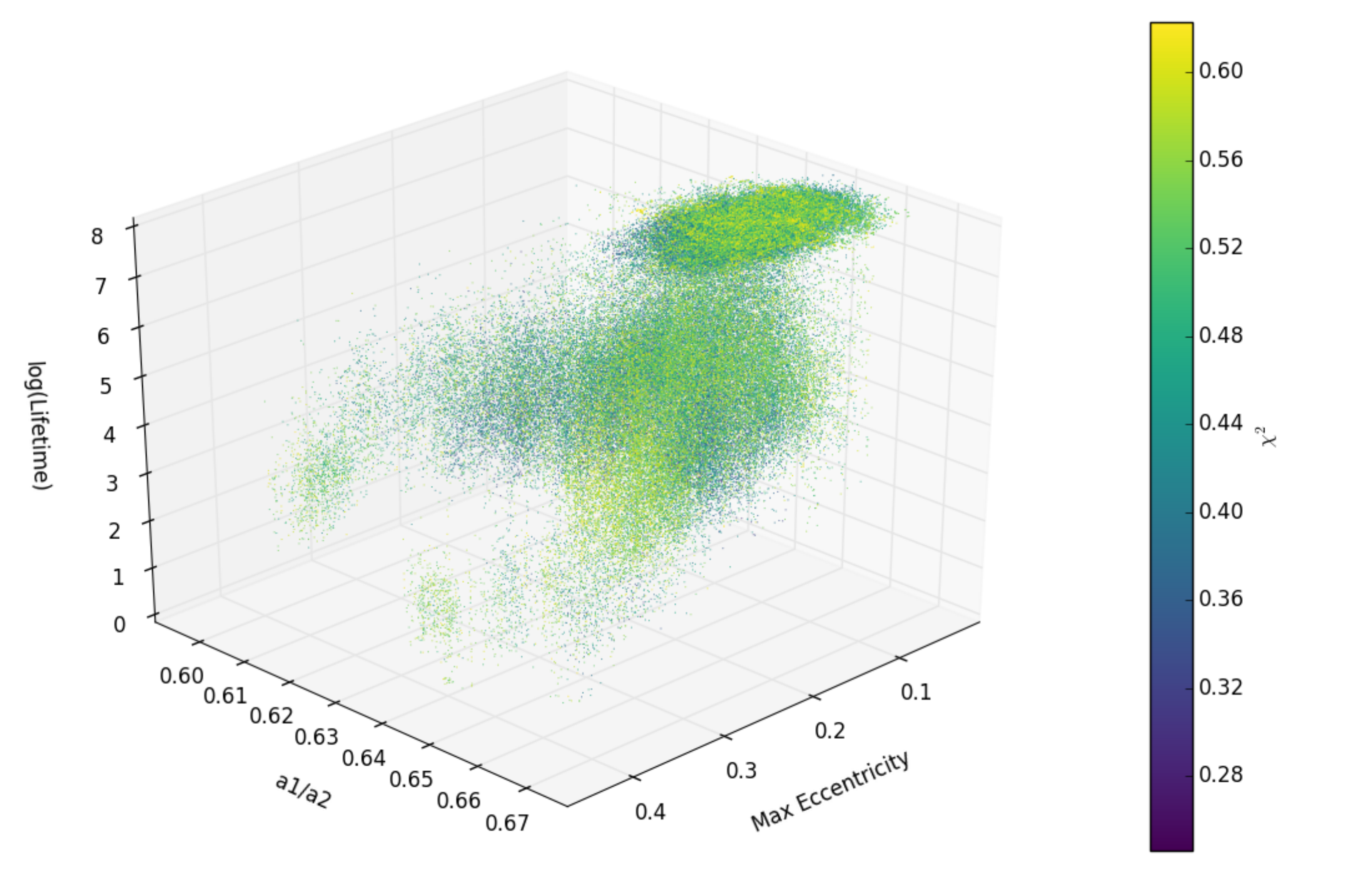}
    \includegraphics[width=0.5\textwidth]{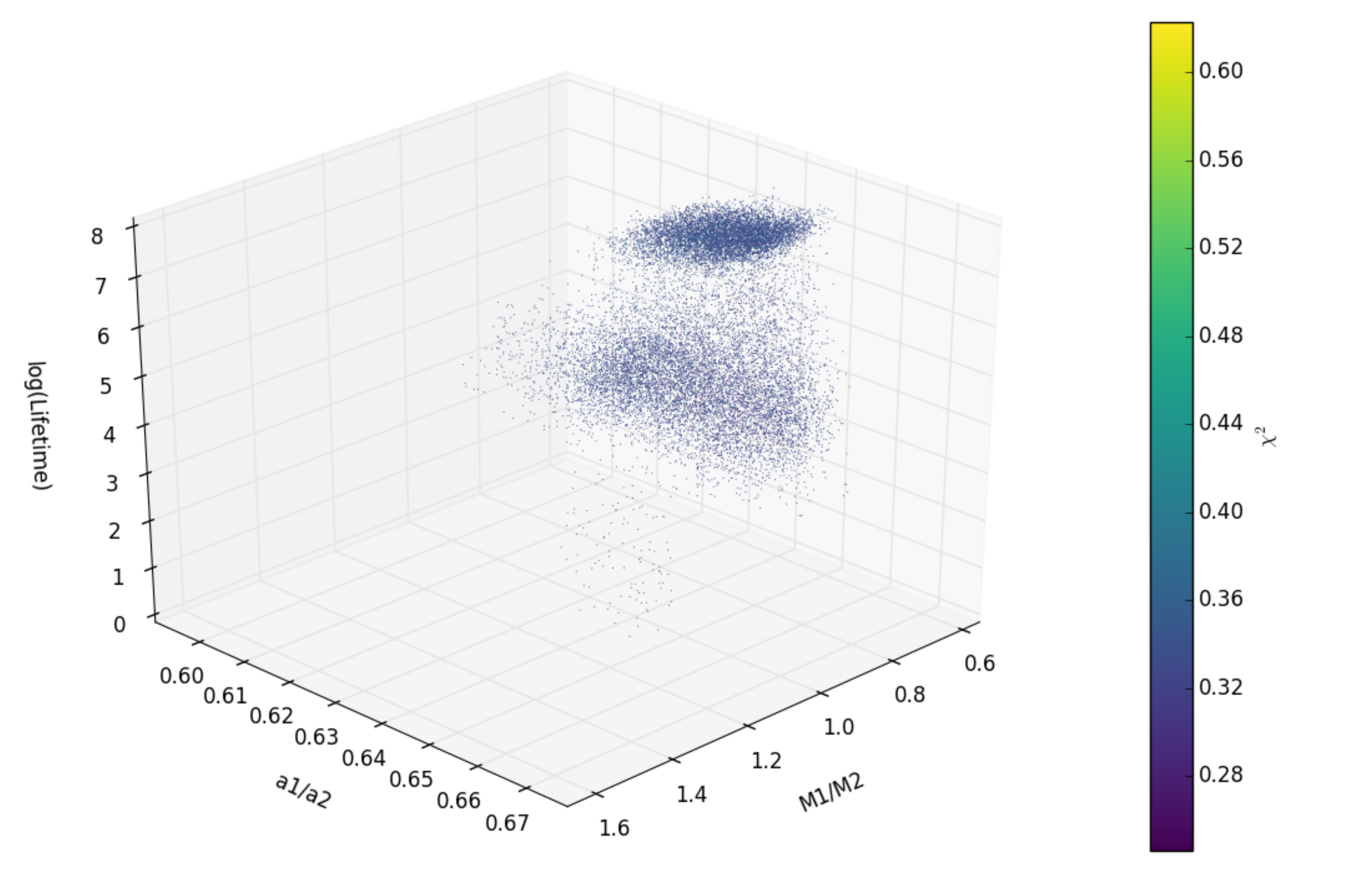}
	\includegraphics[width=0.5\textwidth]{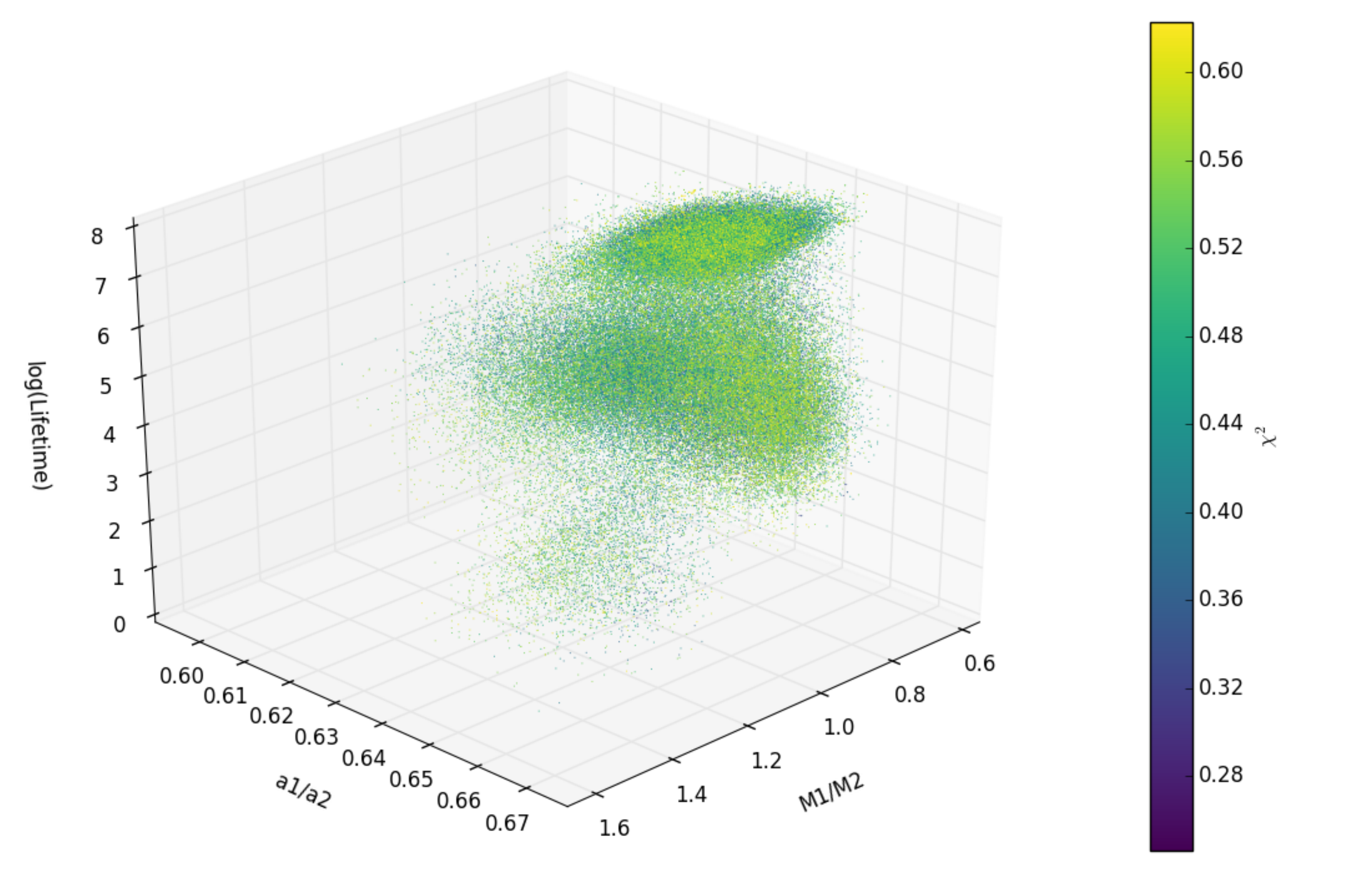}
\caption{The dynamical stability of our revised solution for HD~47366's planets, as a function of the largest initial eccentricity fit to HD47366b and c, and the ratio of their orbital semi-major axes. The colourbar shows the goodness of fit of each solution tested, with the left plot showing only those results within 1-$\sigma$ of the best-fit case, and the right plot showing all solutions tested that fell within 3-$\sigma$ of that scenario. In the online version of this paper, the same plot is available in animated format. {The animated figure lasts 40~s, and shows the 1-$\sigma$ and 3-$\sigma$ distribution of points (equivalent to left and right panels on each line) from a changing perspective rotating around the $z$-axis (log(Lifetime)). These animations help illustrate the regions of parameter space that are more dynamically stable}. \label{RobPairs}}
\end{figure*}

The results of our simulations of planet-pairs around the best-fit solution are shown in Figure~\ref{RobPairs}. It is immediately apparent that a far greater number of tested two-planet scenarios prove dynamically stable in the simulations compared to those based on the S16 solution (shown in Figure~\ref{SatoPairs}). Again, the stable solutions cluster towards lower maximum eccentricities - but the stable region now extends to markedly higher eccentricities. A broad island of stability is clearly visible at eccentricities less than $\sim 0.2$, and for semi-major axis ratios between $\sim 0.63$ and $\sim 0.66$.  

In $a_1 \over a_2$ space, the 2:1 mean motion resonance would be centred on a value of 0.63, with values greater than this revealing pairs of orbits whose periods are more similar to one another. Even values of $a_1 \over a_2$ of 0.66 are still very close to the centre of the 2:1 mean motion resonance - indeed, such orbits would exhibit a period ratio of approximately 13:7 (or 1.86:1) - well within the breadth of the influence of the 2:1 mean-motion resonance. In other words, it seems likely that the stable solutions resulting from our new analysis are facilitated by the influence of that resonance - which would also explain how stable orbits can be maintained up to moderately large orbital eccentricities.  

\subsection{Computation of dynamical MEGNO maps}

In an attempt to further understand the dynamics of the two planets we have 
applied the MEGNO\footnote{Mean Exponential Growth factor of Nearby Orbits}
technique \cite{2000CinSim, 2001Godz, 2003Cincotta} for the numerical assessment 
of chaotic/quasi-periodic orbits in a multi-body dynamical system and has found 
wide-spread applications within the astro-dynamics community \citep{2001Godz,2010Hinse,2014Hinse,2016Contro,2017Wood}. The MEGNO technique 
is especially useful to detect the location of orbital resonances. In summary, MEGNO 
(often denoted as $\langle Y\rangle$) quantitatively measures the degree of stochastic 
behaviour of a non-linear dynamical system. Following the definition of MEGNO 
\citep{2000CinSim} a dynamical system that evolves quasi-periodically the quantity $\langle Y\rangle$ will asymptotically approach the value of 2.0 for $t \rightarrow \infty$. In case of quasi-periodicity, the orbital elements for a given body are bounded and their time evolution described by a few number of characteristic frequencies. For a chaotic time evolution the quantity $\langle Y\rangle$ will diverge away from 2.0. An important point to make is that quasi-periodicity or regular dynamics can only be demonstrated numerically up to the considered integration time. 

Our results were obtained by using a modified version of the fortran-based {\sc $\mu$Farm}\footnote{https://bitbucket.org/chdianthus/microfarm/src
} code \citep{2001Godz, 2003Godz, 2008Godz}. The package utilizes OpenMPI\footnote{https://www.open-mpi.org
} and is capable to spawn a large number of single-task parallel jobs on a given super-computing facility. The package main functionality is the computation of MEGNO over a grid of initial values in orbital elements for a $n$-body problem. The equations of motion and associated variational equations are solved using and effective Gragg-Bulirsh-Stoer {\sc ODEX}\footnote{https://www.unige.ch/$\sim$hairer/prog/nonstiff/odex.f} extrapolation algorithm with step-size control \citep{hairer1993}.

The choice of initial conditions is identical as described in Sec.~\ref{sec:dyn_horner_1}. For a given $(a_c, e_c)$ grid-point of the outer planet a sub-set of various $\omega-M$ parameter combinations is considered within their calculated $1\sigma$ uncertainty. We refer to \cite{2014GodzMiga} for a similar approach in their section 4.5. As for the dynamical study in the previous section we fixed the inner planet to its best-fit parameters. The host star mass was set to $2.19~M_{\odot}$. The single grid-point maximum integration time was set to $10^6$ years due to limited computation resources available. {The MEGNO integration corresponds to over $5\times10^{5}$ orbital periods of the outer planet and hence likely captures the secular time period of the system.}. For each $(a_c,e_c)$ parameter pair we recorded the minimum value of $\langle Y\rangle$ for all tried $\omega_c-M_c$ parameter combinations. The minimum value of $\langle Y\rangle$ is then used to generate a dynamical map over $(a_c,e_c)$ space. This approach ensures that we detect quasi-periodic regions for the probed parameter pairs. However, this approach does not provide information on a specific $\omega_c-M_c$ combination that resulted in a minimum (quasi-periodic) value of 
$\langle Y\rangle$.

\begin{figure}
\subfigure{\includegraphics[width=0.47\textwidth,,trim={4cm 10cm 4cm 10cm},clip]{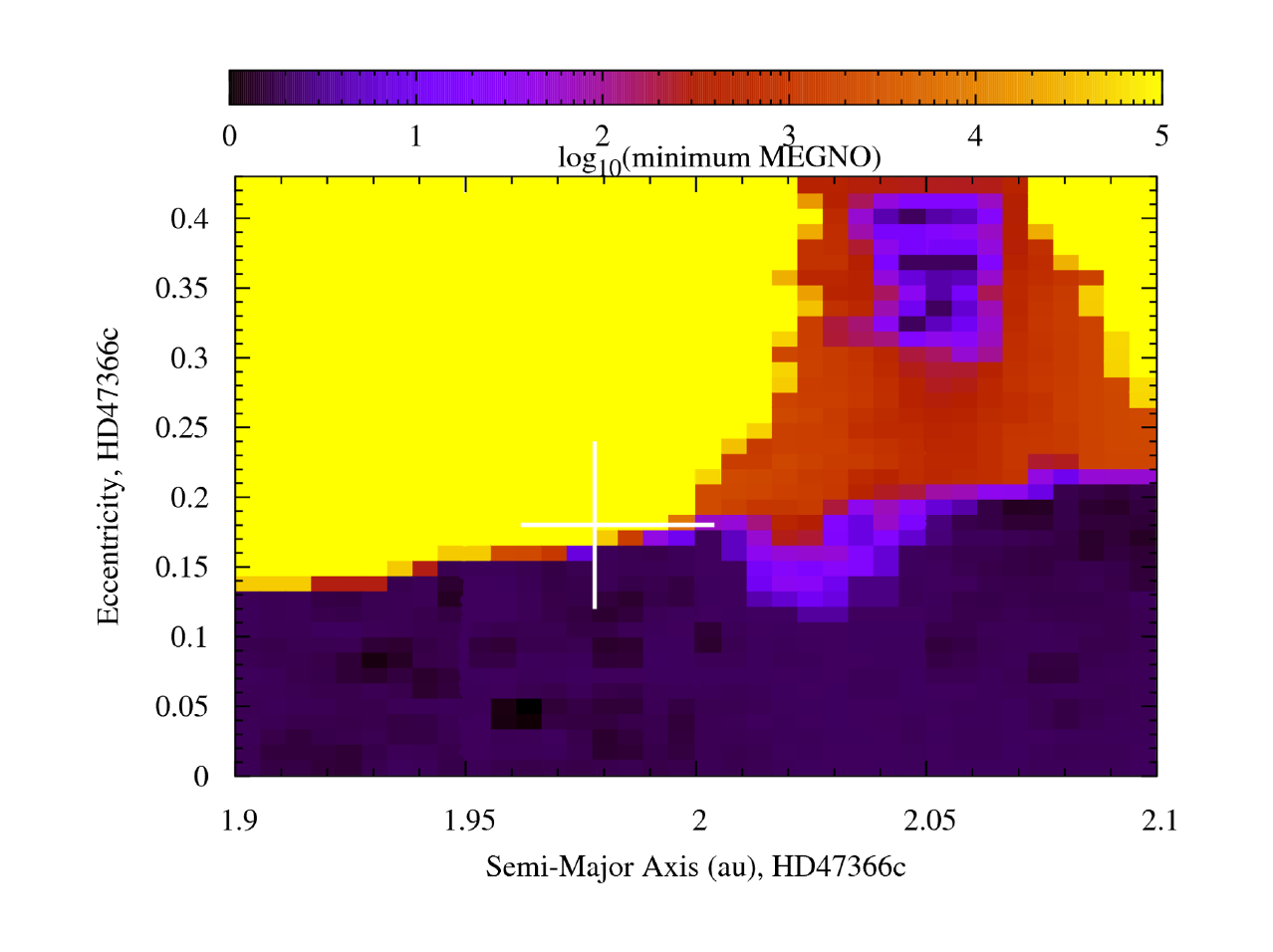}}
\caption{Dynamical MEGNO map considering the $(a_c, e_c)$ space for the outer planet HD~47366c based on the LM + MCMC modelling work. We plot the minimum value of MEGNO. {Quasi-periodic orbits have $\log_{10} Y_{\rm min} ~ 0.3$}. See text for details. Horizontal and vertical bars indicate uncertainty range for the $(a_c,e_c)$ parameters. The inner planet is fixed to best-fit values.}
\label{fig:hd47366_mfarm_map002}
\end{figure}

We present our results in Fig.~\ref{fig:hd47366_mfarm_map002}. The $(a_c,e_c)$ map to a large degree agrees with the life-time map shown in Fig.~\ref{fig:RobContext}. The two independent results complement each other and provide confidence in the numerical results obtained. In overall the considered $(a_c,e_c)$ region is characterized by three areas. i) an area of general orbital instability mainly in the upper left corner ii) an area of stability for low-$e_c$ orbits and iii) an area characterized by an intermediate stability/instability. We point out that the computed MEGNO map considers a somewhat larger range in $(a_c,e_c)$ space as compared to Fig.~\ref{fig:RobContext}. The newly determined LM + MCMC best-fit places the outer planet on the transition region between quasi-periodic (stable) and chaotic dynamics. Long-term orbital stability is still ensured considering the $(a_c,e_c)$ parameter uncertainty range for the outer planet. Low eccentric orbits are preferred prolonging the system's life-time.

\subsection{2:1 near-resonant dynamics}

\begin{figure*}
\centering
\subfigure{\includegraphics[width=0.48\textwidth,trim={4cm 10cm 4cm 10cm},clip]{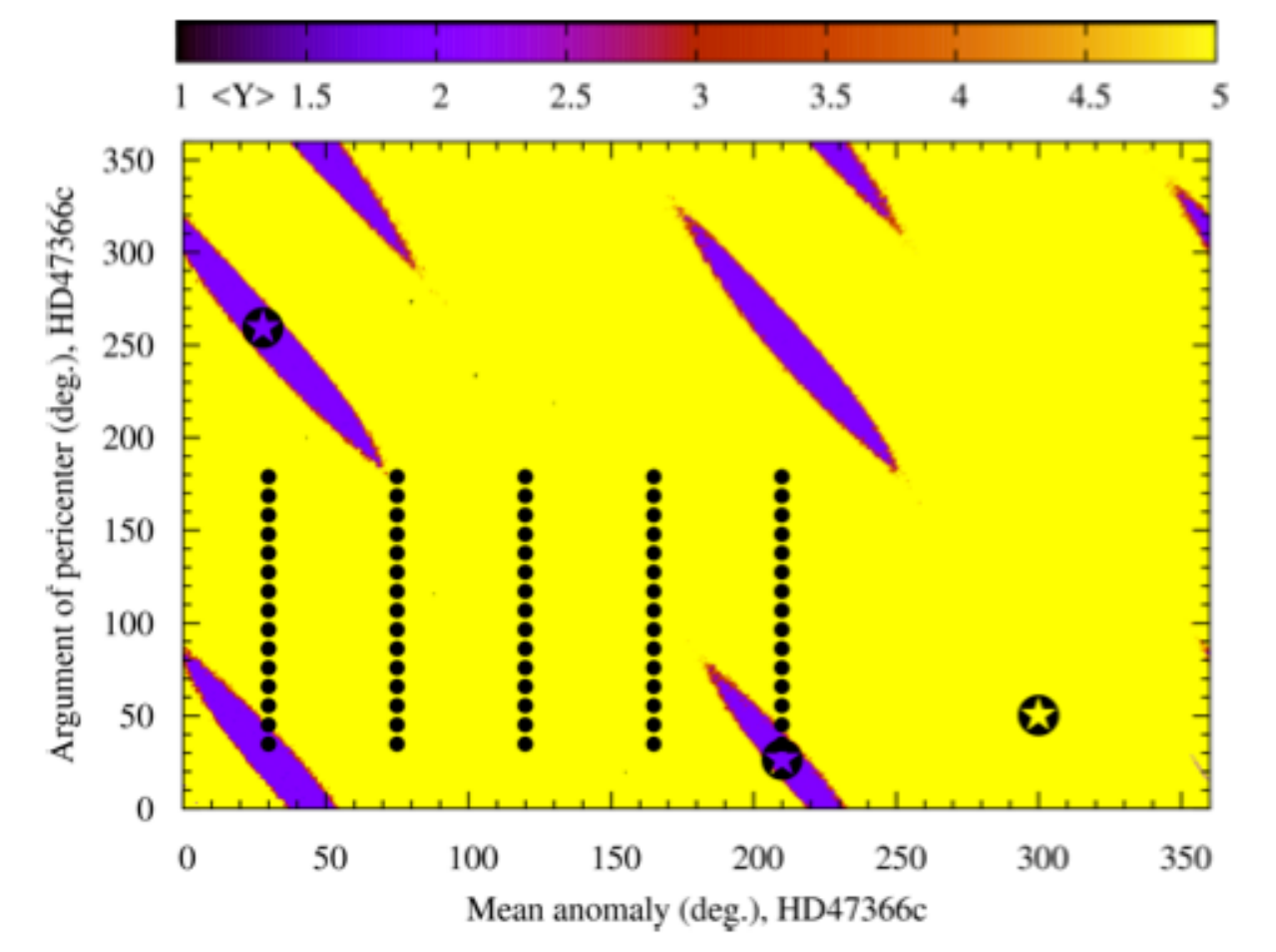}}
\subfigure{\includegraphics[width=0.50\textwidth,trim={4cm 10cm 4cm 9cm},clip]{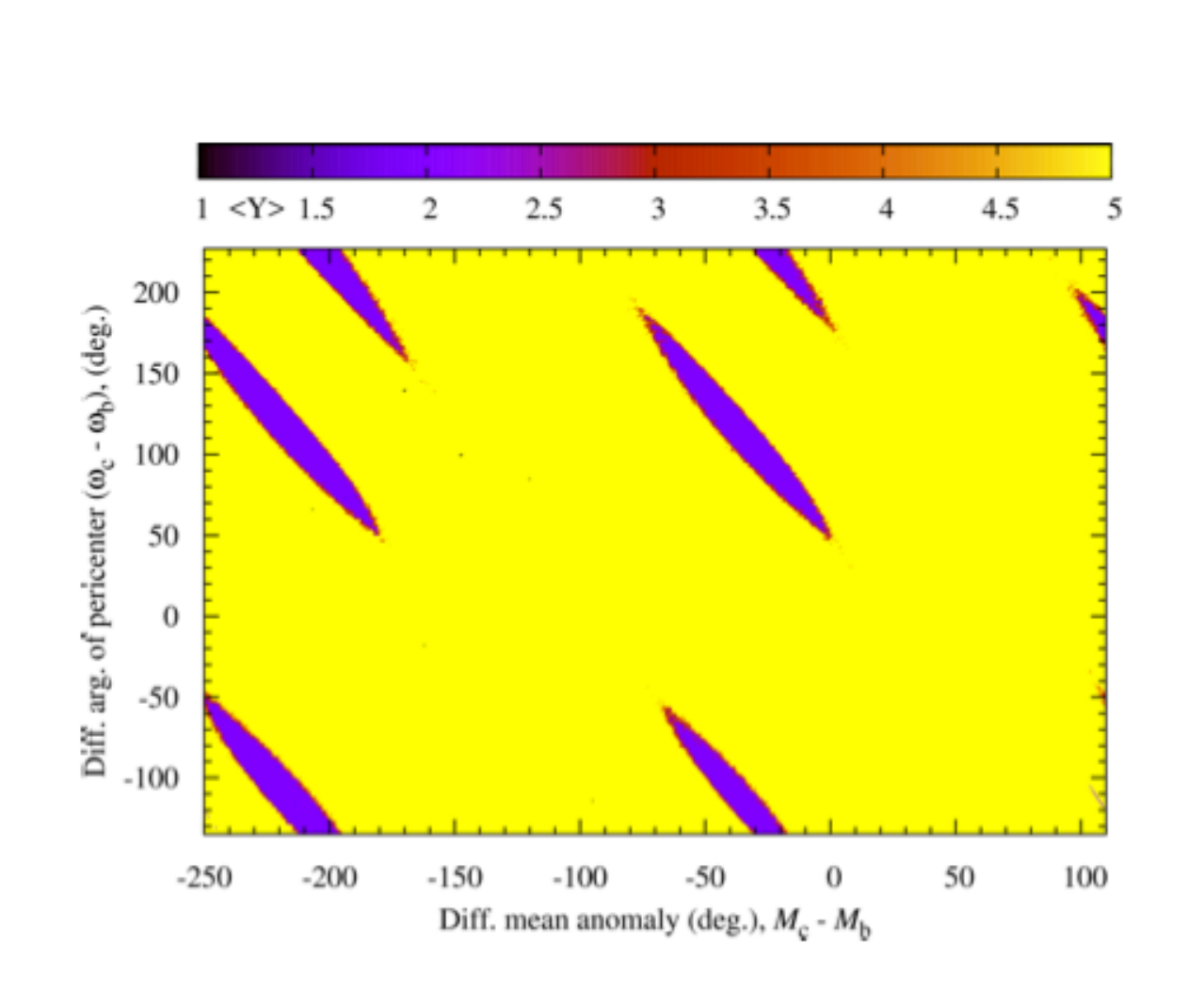}}
\caption{Dynamical MEGNO maps considering the $(M_c, \omega_c)$ space of the outer planet for a fixed $(a_c,e_c)=(2.05\,\rm au,0.367)$. The top panel shows the chosen initial $(M_c,\omega_c)$ parameter range. Black star-like symbols are at $(M_c,\omega_c)=(210^{\circ},26^{\circ})$, $(M_c,\omega_c)=(28^{\circ},259^{\circ})$ and $(M_c,\omega_c)=(300^{\circ},50^{\circ})$. The black dots indicate our probed $a-e-\omega-M$ hypercube. The bottom panel shows the initial conditions relative to the (best-fit) initial $(M_b,\omega_b)=(250^{\circ},134^{\circ})$ of the inner planet.}
\label{fig:hd47366_mfarm_map003}
\end{figure*}

We point out a fourth characteristic in Fig.~\ref{fig:hd47366_mfarm_map002}. The region around $(a_c,e_c)=(2.05~\textnormal{au},0.35)$ exhibits quasi-periodic orbits for some of the probed $\omega_c-M_c$ parameter combinations (shown as black dots in Fig.~\ref{fig:hd47366_mfarm_map003}). We find that the overall dynamics of the region is characterized by the 2:1 mean-motion resonance. The results from our dynamical analysis therefore point toward a two-planet system in a near-resonant orbital architecture. To further characterize the nature of this resonance we have calculated a dynamical MEGNO map over the space $(M_c, \omega_c)$ for a fixed $(a_c,e_c)=(2.05\,\rm au, 0.367)$. The result is shown in Fig.~\ref{fig:hd47366_mfarm_map003}. A total of four stable islands are found. Each corresponds to a particular initial orbital geometry of the two planets resulting in quasi-periodic dynamics. A stable system in 2:1 resonance (for $(a_c,e_c)=(2.05\,\rm au, 0.367)$) is achieved for particular initial differences in apsidal lines ($\omega_c-\omega_b$) and phases ($M_{c}-M_{b}$). The four islands correspond to four initial configurations resulting in a stable 2:1 resonance. In fact, some of these islands were encountered when calculating Fig.~\ref{fig:hd47366_mfarm_map002} as part of the $\omega_c-M_c$ hypercube parameter scan for a chosen $(a_c, e_c)$ pair. This approach is particular effective in identifying orbital resonances.

\begin{figure*}
\centering
\subfigure{\includegraphics[width=0.32\textwidth,trim={5cm 10cm 5cm 10cm},clip]{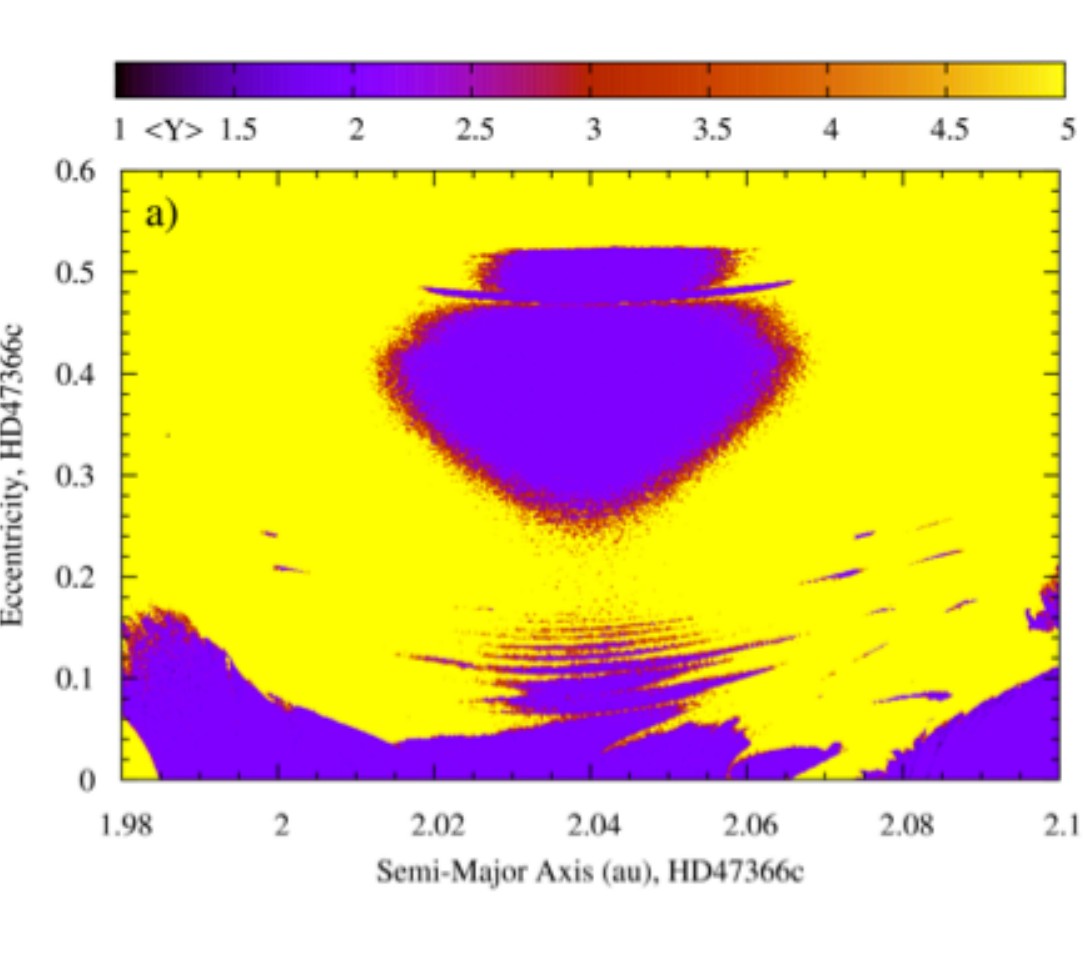}}
\subfigure{\includegraphics[width=0.32\textwidth,trim={5cm 10cm 5cm 10cm},clip]{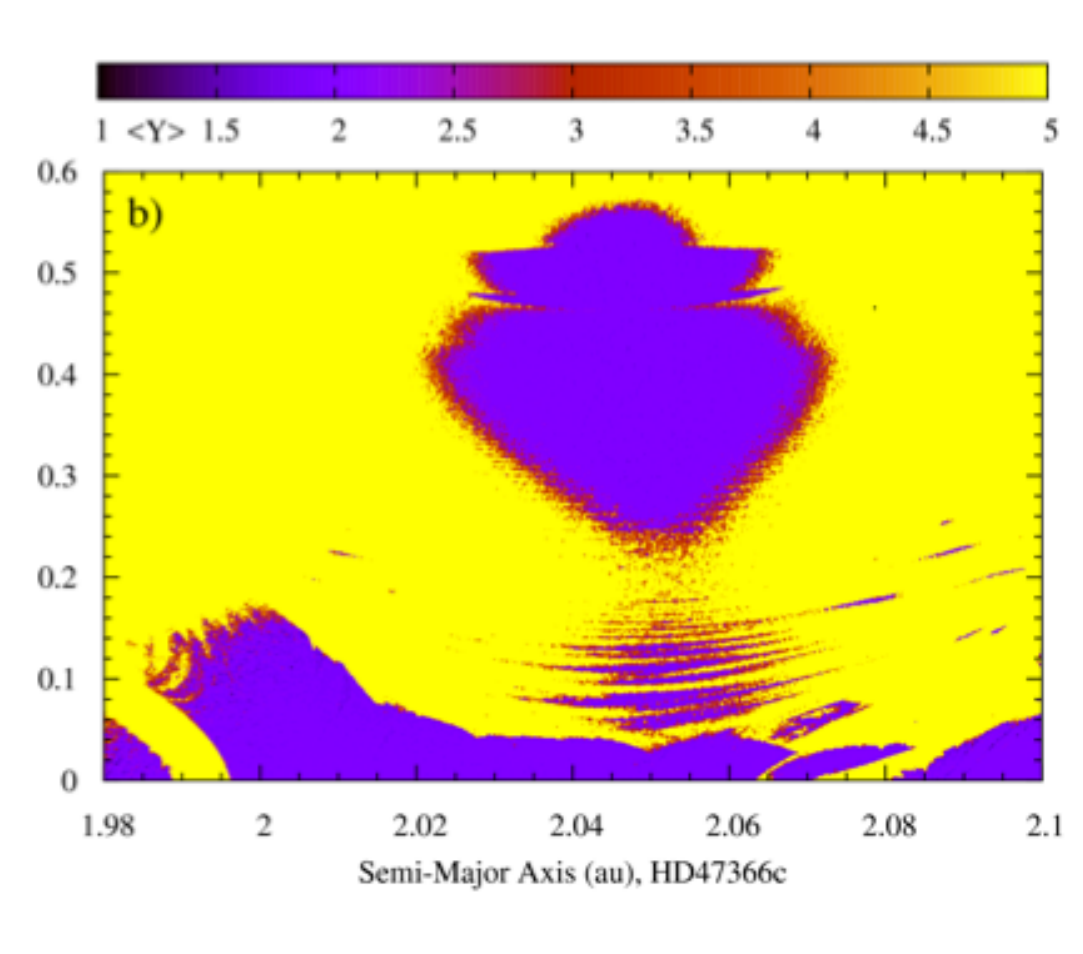}}
\subfigure{\includegraphics[width=0.32\textwidth,trim={5cm 10cm 5cm 10cm},clip]{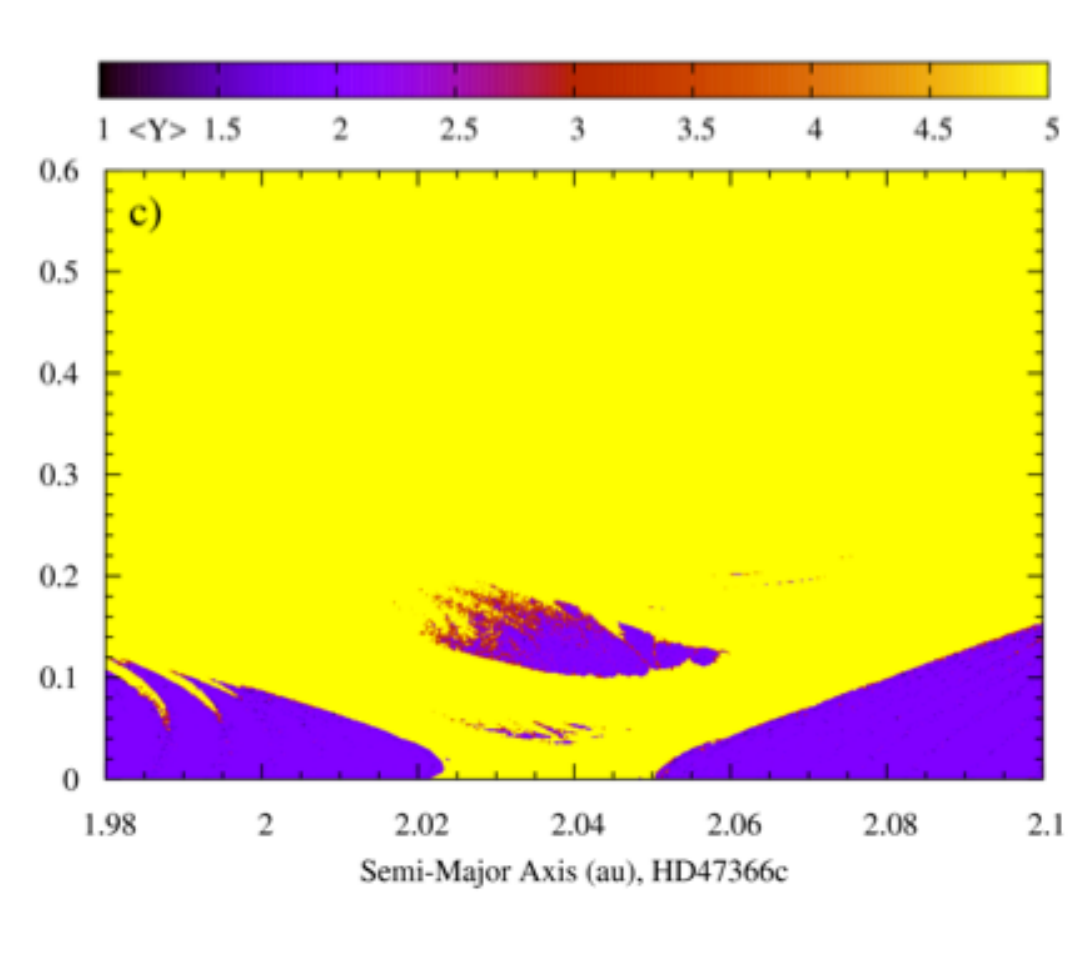}}
\caption{Dynamical MEGNO maps for three different choices in initial mean anomaly and argument of pericenter for the outer planet. The three orbits are indicated by a black dot in Fig.~\ref{fig:hd47366_mfarm_map003}. \emph{Left panel:} $(M_c,\omega_c)=(210^{\circ},26^{\circ})$. \emph{Middle panel:} $(M_c,\omega_c)=(28^{\circ},259^{\circ})$. \emph{Right panel:} $(M_c,\omega_c)=(300^{\circ},50^{\circ})$.}
\label{fig:hd47366_mfarm_map005_map006_map009}
\end{figure*}

We have repeated the calculation of Fig.~\ref{fig:hd47366_mfarm_map002} for particular $(M_c,\omega_c)$ combinations corresponding to the approximate center of two libration islands in Fig.~\ref{fig:hd47366_mfarm_map003} (shown as black star-like symbols). One particular $(M_c,\omega_c)=(300^{\circ},50^{\circ})$ pair is deliberately chosen to be within the chaotic region for comparison. The corresponding dynamical MEGNO maps are shown in Fig.~\ref{fig:hd47366_mfarm_map005_map006_map009}. In comparison with Fig.~\ref{fig:hd47366_mfarm_map002} we now clearly identify the quasi-periodic island associated with the 2:1 resonance for two chosen initial configurations of apsidal line and phase differences. For the initial condition chosen from the chaotic region we find that the stability island now disappears as expected.

\begin{figure}
\centering
\subfigure{\includegraphics[width=0.45\textwidth,trim={0cm 7cm 0cm 8cm},clip]{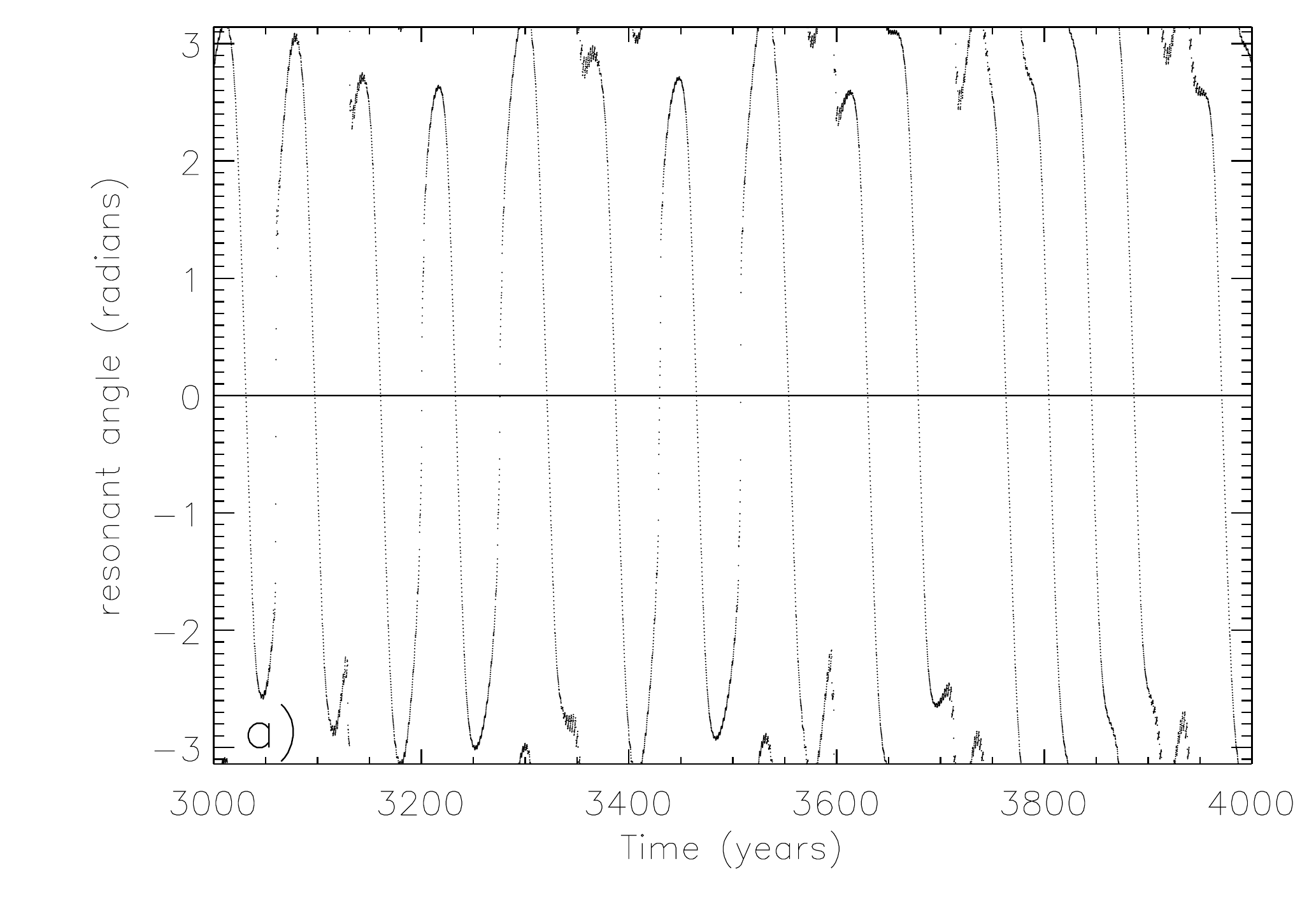}}
\subfigure{\includegraphics[width=0.45\textwidth,trim={0cm 7cm 0cm 8cm},clip]{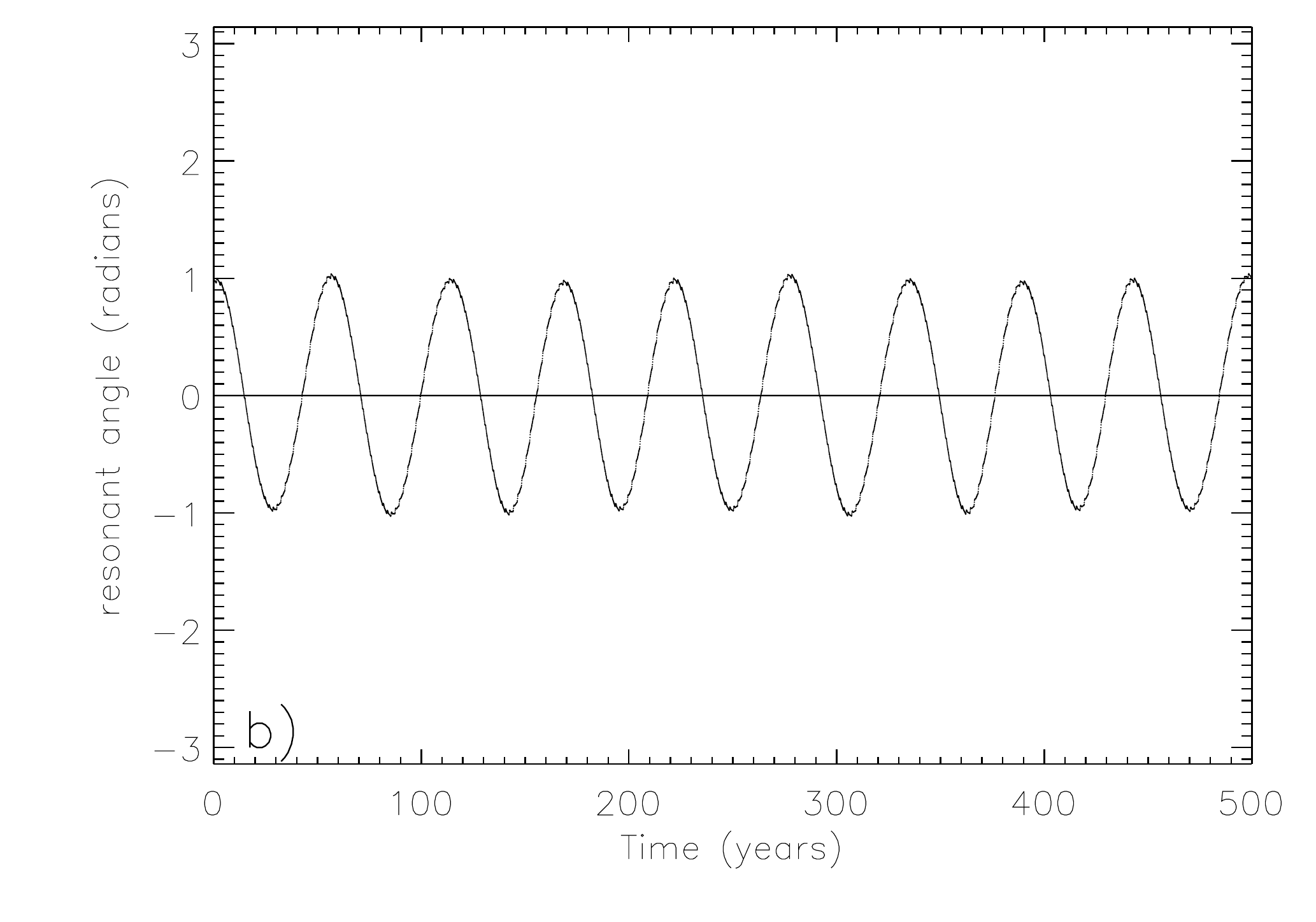}}
\caption{Time evolution of the critical resonance angle. In the top panel the angle is librating demonstrating locking in 2:1 mean-motion resonance. The bottom panel demonstrates alternation between librations and circulations.}
\label{fig:hd47366_resangle}
\end{figure}

As a last exercise we have checked the time evolution of the critical resonant angle for two particular initial conditions. The resonant angle for the 2:1 mean-motion resonance is

\begin{equation}
    \phi = 2\lambda_c - 1\lambda_b - \omega_b
\end{equation}
\noindent
where $\lambda$ is the mean longitude for either planet. In Fig.~\ref{fig:hd47366_resangle} we plot this angle for the two cases: Panel a: $(a_c,e_c,\omega_c,M_c)=(2.05\,\rm au, 0.01,259^{\circ},28^{\circ})$. Panel b: $(a_c,e_c,\omega_c,M_c)=(2.05\,\rm au, 0.01,50^{\circ},300^{\circ})$. From direct numerical integrations using the {\sc Mercury} \citep{1999Chambers} integrator we find the quasi-periodic and chaotic time evolution of the resonant angle for the two chosen initial conditions demonstrating stable and chaotic dynamics.

\section{Discussion}
\label{sec:dis}

In recent years, a number of studies have shown the importance of performing dynamical tests of newly proposed planetary systems, in order to verify that the proposed systems are truly dynamically feasible \citep[e.g.][]{2011Horner,2014Wittenmyer,2017Horner}. In this light, we performed a detailed dynamical analysis of the proposed HD~47366 planetary system, as detailed in \cite{2016Sato}. Our results indicated a good chance that the two-planet system proposed in that work would be dynamically unstable, and therefore raises its plausibility into question. Since the observational data clearly show evidence for two strong signals, we chose to perform a fresh analysis of that data, in order to determine whether a dynamically stable solution could be found that is an adequate fit to the observations. 

To fully characterise the proposed system, we examined a binned data set of radial velocity observations of the two planet HD~47366 system using two independent methods to determine the best-fit parameters of the system, under the assumption of co-planar orbits. We find that the system properties determined by each method are generally consistent. For HD~47366 c, all the orbital parameters agree within uncertainties. For HD~47366 b the orbital angles $\omega$ and $M$ determined by each method are not consistent with each other. However, the large uncertainties on those values ($> 90\degr$) makes the significance of this discrepancy small ($< 3-\sigma$) and we conclude that the results of the separate analyses are therefore in agreement.

Once we had obtained our new solution for the two-planet system, we performed a fresh dynamical analysis, examining whether our solution offered greater prospects for stability than that proposed in S16. In strong contrast to that earlier work, we found that our new solution resulted in a large number of potentially stable scenarios, all of which offered an excellent fit to the observational data. Those solutions nestled close to the location of the mutual 2:1 mean-motion resonance between the two proposed planets. This enhanced stability affords us greater confidence in the validity of our new solution. 

The orbital parameters of the best fitting model produce an architecture with the outer planet having an orbital period $\sim$ 1.9 times that of the inner planet. Multi-planet systems discovered by \textit{Kepler} exhibit a significant pile-up at orbital periods close to, but outside of, the 3:2 and 2:1 resonances with an inner planet \citep{2014WangJi}, typically inferred to be a relic of planetary migration in the presence of an accreting circumstellar disc. Given the proximity of our new solution to the 2:1 mean-motion resonance, it seems plausible that the HD~47366 planetary system can be added to the catalogue of such compact planetary systems.

As HD~47366 evolves off the main sequence and undergoes significant mass loss, its planetary system will be destabilised. The cause of this destabilisation are the tides raised by the planets on the puffed up star. Tidal forces act to damp the planet's semi-major axis and eccentricity. The two possible outcomes of this process are either engulfment by the host star or evaporation into interstellar space, depending on the initial semi-major axis and mass of the planetary companion \citep{2012Mustill}. Modelling has shown that tidal effects become important when the periastron distance of the planet approaches between two and three stellar radii. In the case of HD~47366, both planets have semi-major axes small enough that they will likely be subject to tidal forces as the host star evolves. HD~47366's planets will therefore undergo rapid orbital decay and be engulfed during its post-main sequence evolution, similar to the expected fate of the eccentric gas giant around HD~76920 \citep{2017bWittenmyer}.

\section{Conclusions}
\label{sec:con}

We have performed a thorough dynamical reanalysis of the planetary system proposed to orbit the star HD~47366 in S16. Our simulations cast doubt on the dynamical feasibility of the solution proposed in that work. As a result, we have performed a detailed reanalysis of the available observational data for the system, and have used that to produce an improved solution for the proposed two-planet system.

Through our Levenberg-Marquardt reanalysis of the two-planet system proposed 
around HD~47366, we have demonstrated a low(er) eccentricity orbital solution 
exists compared to that proposed by S16. This solution is shown to be 
dynamically stable for periods up to 100~Myr, and is comparable to the 
orbital parameters derived by our Bayesian reanalysis of the system. 

We present this work as a cautionary tale in exoplanet dynamics -- the 
best-fit solution derived from Keplerian modelling of radial velocities 
may only be a local one, particularly if the resulting solution requires 
contrived architectures to be stable over periods comparable to the 
lifetime of the host star. By expanding the parameter space explored in 
the fitting process we have determined a much more dynamically 
plausible solution for the architecture of the HD~47366 system.

\acknowledgements

{The authors thank the anonymous referee for their constructive criticism.}

This research has made use of the SIMBAD database, operated at CDS, Strasbourg, France. 

This research has made use of NASA's Astrophysics Data System.

This research has been supported by the Ministry of Science and Technology of Taiwan under grants MOST104-2628-M-001-004-MY3 and MOST107-2119-M-001-031-MY3, and Academia Sinica under grant AS-IA-106-M03.

{\textit{Software:} {\sc Systemic} \citep{2009Meschiari}, {\sc astroemperor} (https://github.com/ReddTea/astroEMPEROR), {\sc matplotlib} \citep{2007Hunter}, {\sc Mercury} \citep{1999Chambers}, {\sc emcee} \citep{2013ForemanMackey}.}

\bibliographystyle{apj}
\bibliography{hd47366_refs}

\end{document}